# Pushing the limits of flow strength in diamond

Anirudh Hari[1,2,3]†, Kento Katagiri[1,2,3,4,5]†*, Wanghui Li[6]†, Dorian P. Luccioni[1,2,3], Rayen Lin[1,3,7], Sophie E. Parsons[1,2,3], Zipeng Xu[1,2,3], Rohit Hari[8], Tharun Reddy[1,2,3], Ernest W. Cubit II[1,2,3], Alexis Amouretti[4], Jon H. Eggert[9], Yuichi Inubushi[10,11], Tetsuo Irifune[12,13], Sara J. Irvine[2,3,14], Ryosuke Kodama[4,5], Michel Koenig[4,15], Laura Madril[1,2,3], Takeshi Matsuoka[16], Kohei Miyanishi[11], Hirotaka Nakamura[4], Norimasa Nishiyama[17], Takuo Okuchi[18], Masato Ota[19], Toshimori Sekine[4,20], Yusuke Seto[21], Toru Shinmei[12], Keiichi Sueda[11], Yoshinori Tange[10], Sota Takagi[22], Tadashi Togashi[10,11], Yuhei Umeda[18], Yifan Wang[1,2,3], Makina Yabashi[10,11], Toshinori Yabuuchi[10,11], Norimasa Ozaki[4,5], and Leora E. Dresselhaus-Marais[1,2,3]*

[1]Stanford University, Department of Materials Science and Engineering, California 94305, USA

[2]SLAC National Accelerator Laboratory, California 94025, USA

[3]Stanford University, PULSE Institute, California 94305, USA

[4]Graduate School of Engineering, Osaka University, Osaka 565-0871, Japan

[5]Institute of Laser Engineering, Osaka University, Osaka 565-0871, Japan

[6]Institute of High Performance Computing (IHPC), Agency for Science, Technology and Research (A*STAR), Singapore 138632, Republic of Singapore

[7]University of California, Los Angeles, Department of Materials Science and Engineering, California 90095, USA

[8]Georgia Institute of Technology, Department of Electrical and Computer Engineering, Georgia 30332, USA

[9]Lawrence Livermore National Laboratory, California 94550, USA

[10]Japan Synchrotron Radiation Research Institute, Hyogo 679-5198, Japan

[11]RIKEN SPring-8 Center, Hyogo 679-5148, Japan

[12]Geodynamics Research Center, Ehime University, Ehime 790-0826, Japan

[13]Earth-Life Science Institute, Tokyo Institute of Technology, Tokyo 145-0061, Japan

[14]Stanford University, Department of Applied Physics, California 94305, USA

[15]LULI, CNRS, CEA, École Polytechnique, UPMC, Univ Paris 06: Sorbonne Universités, Institut Polytechnique de Paris, 91128 Palaiseau cedex, France

[16]Open and Transdisciplinary Research Initiatives, Osaka University, Osaka 565-0871, Japan

[17]Laboratory for Materials and Structures, Tokyo Institute of Technology, Kanagawa 226-8503, Japan

[18]Institute for Integrated Radiation and Nuclear Science, Kyoto University; Osaka 590-0494, Japan

[19]National Institute for Fusion Science, Gifu, Japan

[20]Center for High-Pressure Science and Technology Advanced Research, Shanghai 201203, China

[21]Graduate School of Science, Kobe University; Hyogo 657-0013, Japan

[22]Earth and Planets Laboratory, Carnegie Institution for Science, Washington DC 20015, USA

†These authors contributed equally to this work.

Correspondence: kentok@stanford.edu, leoradm@stanford.edu

## Abstract

Extreme pressures and temperatures create conditions that allow even hard and brittle materials to deform plastically. Despite extensive research, the upper limits of flow strength, the resistance to plastic flow, remain uncertain, and the mechanisms driving deformation at the relevant stresses are a subject of debate. Using femtosecond *in situ* X-ray diffraction experiments and large-scale molecular dynamics simulations, we demonstrate that stacking fault-mediated strengthening enables shock-compressed nano-polycrystalline diamond to achieve a peak flow strength of 107 ± 5 GPa at a stress of 227 ± 8 GPa. Our findings show that extreme conditions can unlock unusual strength via mechanisms that can be used as design tools in targeted applications.

## Introduction

The flow strength represents a material's resistance to plastic deformation when stressed beyond its elastic limit (*1*). This key mechanical property governs wide-ranging phenomena including the failure of structural components (*2*), the resilience of spacecraft shielding (*3*) and the behavior of planetary interiors under extreme gravitational compression (*4*). To design stronger materials for extreme environments, it is essential to identify the mechanisms of flow strength and the stress regimes in which they operate.

As the strongest natural material with phase stability at extreme stresses, diamond is the ideal material to explore the limits of flow strength and its mechanisms. Under ambient conditions, diamond fractures when stressed beyond its elastic limit. However, extreme pressures and temperatures can create conditions that suppress cracking and mobilize defects, allowing even brittle materials like diamond to deform plastically and exhibit flow strength (*5-7*).

At the most extreme pressures and temperatures, materials may either weaken or strengthen depending on the active deformation mechanisms (*8-11*). Under pressures of hundreds of gigapascals (GPa), wave-profile analyses have sought to infer the flow strength in shocked single-crystalline diamond, but large uncertainties in the measurements and models have hindered interpretation of the results (*11*). One *in situ* XRD study on shocked micro-polycrystalline diamond found some evidence of strength, but the small sampling of diffracting grains similarly led to large uncertainties that obscure any clear trends (*12*).

Meanwhile, recent X-ray images of single-crystalline diamond shock-compressed up to 184 ± 16 GPa revealed crisscrossing diagonal bands along the {111} planes (*13*). These features were attributed to bundles of stacking faults bound by partial dislocations generated by the shock. While these measurements suggest partial dislocations as a possible deformation mechanism, what remains missing is a connection between pressure-induced defects and diamond's flow strength. This connection is key to developing predictive strength models at extreme conditions.

Full-density nano-polycrystalline diamond (NPD) is among the strongest forms of diamond (*14, 15*), reaching an extremely high ultimate flow strength of up to 23 GPa under quasi-static compression (*7*). Therefore, we chose NPD as an ideal material to study novel strengthening behavior under extreme conditions. In this work, we used *in situ* XRD to measure the flow strength and microstructural response of shock-compressed NPD. Our results show that NPD reaches an unprecedented peak flow strength when compressed to 227 ± 8 GPa. At higher stresses, NPD loses its strength develops a weak <110> fiber texture that is characteristic of dislocation plasticity. Using large-scale molecular dynamics (MD) simulations, we reveal that the flow strength is governed by a complex array of competing strengthening and softening mechanisms.

## Results

Ultra-high flow strength

Using uniaxial shock loading with a high-intensity optical laser ($\sim 10^{12}$ W/cm²), we generated multi-megabar pressure conditions far beyond what was achieved with quasi-static loading. We probed the flow strength of the shocked NPD using femtosecond *in situ* XRD at the SPring-8 Angstrom Compact Free Electron Laser (SACLA) (*16*) (Fig. 1A). The compressed states were probed ~1 ns after shock entry into NPD using an ultrafast (<10 fs) and intense ($\sim 10^{11}$ photons/pulse) XFEL pulse (*17*), and diffraction patterns were captured on flat-panel pixel array detectors. To reveal the underlying deformation mechanisms, we paired these in situ XRD measurements with large-scale MD simulations of NPD shock-compressed to various peak stresses.

We used *in situ* XRD to directly measure the lattice strains in shocked NPD and extract the material's flow strength. A representative diffraction pattern is shown in Fig. 1B. In such patterns, the position of each diffraction ring encodes the spacing between atomic planes in the crystal (*d*-spacing), while intensity variations around the rings reflect how grains are oriented. Line profiles taken at specific azimuthal sectors are shown in Fig. 1C. The diffraction data show three distinct peaks for each reflection, corresponding to uncompressed, elastically compressed, and plastically compressed regions within the sample. In NPD, the longitudinal sound speed (~18 km/s) is much faster than the bulk sound speed (~11 km/s) (*18*), so the material supports a two-wave shock structure—an elastic wave followed by a plastic wave—even up to multi-megabar stresses.

Our diffraction measurements allowed us to determine the flow strengths of NPD shocked to various peak stresses. The azimuthal variations of the *d*-spacing were analyzed to extract the lattice strain components along the shock direction ($\varepsilon^{\ell}_{\parallel}$) and the directions transverse to it ($\varepsilon^{\ell}_{\perp}$). The difference between these components gives the differential lattice strain, which we multiplied by the shear modulus to compute the flow strength $2\tau$. We also calculated flow strength from our MD simulations (Fig. S1), where the longitudinal and transverse stresses were measured directly.

The flow strengths measured from both experiments and simulations are plotted as a function of shock stress and total strain in Fig. 2, and show an increase followed by a reduction with shock intensity. The experimentally measured flow strength reaches a peak value of 107 ± 5 GPa at a shock stress of 227 ± 8 GPa and total strain of 0.262 ± 0.006. At stresses below 227 ± 8 GPa, the flow strength increases with shock stress, while at higher stresses, the strength decreases and becomes negligible at shock stresses of 565 ± 14 GPa in the experiments and 623 GPa in the simulations. Our results identify two deformation regimes characterized by stress hardening followed by softening with increasing shock intensity.

Ultrafast grain reorientation from dislocation plasticity

In crystal lattices, dislocations are line defects that move along specific slip systems to carry plastic deformation (*19*). In diamond, the primary slip systems are along the {111} planes, in the <110> directions. Our XRD experiments provide mechanistic evidence of dislocation-mediated slip through changes in grain orientations, termed crystal texture.

Under compressive loading, plastic deformation mediated by dislocation slip typically leads to the reorientation of crystallites. This occurs because slip within grains alters their shape in specific ways, causing the grains to rotate to meet the lateral confinement requirement imposed by the plane strain shock condition. In single crystals and textured polycrystals, crystal rotation appears as movement of diffraction spots along the azimuthal angle (*20*). This has been used to infer active slip systems and twinning in shocked tantalum, iron, copper and magnesium (*21-24*). However, this approach cannot be used directly in an initially untextured material like NPD.

Instead, we can assess grain reorientation as an average effect through the emergence of crystallographic texture.

Our diffraction measurements show development of a preferred orientation in NPD shocked to 425 ± 14 GPa and above, as evidenced by azimuthal variations in diffraction intensity in the plastically compressed fraction (Fig. 3A). This reorientation of grains is detectable less than one nanosecond after the shockwave enters the sample. Simulated diffraction patterns generated by forward modeling various textures indicate that our results are consistent with the emergence of a weak <110> fiber texture (Fig. 3B). This means the grains reorient such that their <110> directions align with the shock compression axis, while orientations transverse to this axis remain randomly distributed.

By generating simulated two-dimensional XRD patterns from the MD atomic positions at 623 GPa, we confirm consistency with the experimentally observed diffraction patterns at high stresses (Fig. 3C). Inverse pole figures from the MD results before and after compression confirm that NPD with an initially random texture forms a <110> fiber texture upon shock (Fig. 3D).

These results can be interpreted in the context of face-centered cubic (FCC) metals, which share much of their symmetry elements with diamond. When compressed, untextured FCC metals typically develop a <110> fiber texture due to slip along the {111}<110> systems within the grain interiors. This texture formation is caused by the lateral confinement effect described earlier and has been demonstrated both theoretically and experimentally in compressed FCC metals under a variety of loading conditions (*25-29*). Therefore, we interpret the grain reorientation to a <110> fiber texture as evidence of increasing dislocation-mediated slip within the grains at high stresses.

### Competing dislocation plasticity mechanisms

The diamond-cubic structure, which can be represented as a face-centered cubic lattice with a two-atom basis, supports full dislocations with Burgers vectors a/2<110> gliding on the {111} slip planes. These dislocations can dissociate into two partials (a/6<112>{111}), lowering the energy barrier for plastic deformation. The two partial dislocations are separated by a planar defect called a stacking fault.

Our MD simulations reveal that the transition from hardening to softening at 260 GPa coincides with a shift in the predominant dislocation character (Fig. 4A). Across all stresses, both partial and full dislocations are observed (Fig. 4B). However, the trends in dislocation densities between these two dislocation types are different.

Up to 260 GPa, increasing shock stress causes a rise in both partial and full dislocation densities, with higher densities of partial dislocations, as illustrated by a grain-scale visualization at 196 GPa (Fig. 4C). Above 260 GPa, the density of partials slightly reduces and levels off, while the density of full dislocations continues to rise and exceeds that of partials, a shift confirmed by visualizing the same grain at 623 GPa (Fig. 4D). At high stresses, we observe additional shear stress localization at grain boundaries as full dislocations become dominant.

We propose that the hardening behavior below 260 GPa is linked to the increase in partial dislocation density. The stacking faults that are bound by partials constrain dislocation motion to a specific plane, thereby reducing the mean free path of dislocations moving on other planes and leading to stacking fault-mediated strain hardening (*30*). This demonstrates a new strengthening effect in diamond, in addition to the established grain boundary and nano-twin strengthening (*14, 15, 31*). At higher stresses, the substantial increase in full dislocation density likely facilitates easier plastic flow, contributing to both the observed softening and texture formation.

## Discussion

In this work, we report an ultra-high peak flow strength of 107 ± 5 GPa in shock-compressed NPD, representing a five-fold increase over its ultimate flow strength under quasi-static compression (*7*). This peak flow strength is also higher than previous results on hard materials including micro-polycrystalline and single-crystalline diamond (*11, 12*), and is more than ten times that of other strong materials under similar conditions, such as alumina and silicon carbide (*32, 33*) (Fig. 2B). Beyond establishing a new benchmark for flow strength, our results also show clear trends of flow strength with respect to stress and strain that have not been previously demonstrated in shocked diamond (Fig. 2).

### Competing strengthening and softening mechanisms

In shock experiments, high stresses are inherently accompanied by high temperatures. We calculated shock temperature for our experimental measurements by fitting to a high-pressure equation of state (EOS) of carbon (*34*). For our MD measurements, we calculated temperature directly using the two atomic velocity components perpendicular to the shock direction. Fig. 5A shows these temperatures plotted against the shock stress, along with experimental and simulated melt curves (*35, 36*).

We propose that the non-monotonic trend of flow strength with increasing peak stress arises from a combination of competing thermodynamic effects and deformation mechanisms. Based on our MD results and observations from the literature, we identify several possible strengthening and softening mechanisms in shocked NPD.

In the low-stress hardening regime, most of the increase in flow strength with stress is caused by the rise in partial dislocation density and associated stacking faults. Additionally, higher pressures increase atomic packing and strengthen interatomic bonding, thereby raising the shear modulus (*37*) and thus the flow strength. Although the concurrent temperature rise may induce some thermal softening, the temperatures remain relatively low in this regime (<2000 K), so this effect remains small.

We interpret that the shift from hardening to softening above 260 GPa is primarily due to the rise in density of full dislocations (Fig. 4), whose motion accommodates greater plastic strains. The order-of-magnitude increase in full dislocation density in this regime overpowers any remaining hardening effects from stacking faults and the pressure-induced increased shear modulus. The dominance of full dislocations in this regime is likely due to a combination of both higher stresses and temperatures. Higher stresses lead to more dislocations that accommodate the increased plastic strain, and high temperatures further contribute to softening by increasing dislocation mobility.

High temperatures are also known to promote bond-weakening and amorphization, both of which may contribute to loss of strength in the softening regime. MD snapshots of a grain at various shock stresses are shown in Fig. 5B–E. Intermediate stresses in the softening regime feature a modest level of amorphization at the grain boundaries, with 5% disordered atoms at 260 GPa (Fig. 5C) and 9% at 366 GPa (Fig. 5D). At the highest stress of 623 GPa, 30% of the atoms are disordered, still primarily at the grain boundaries (Fig. 5E). This higher level of amorphization is expected as the material approaches the melting point. The significant grain boundary thickening may provide an additional mechanism for softening at the highest stresses via grain boundary-mediated plasticity.

Additional strengthening mechanisms include Hall-Petch grain-boundary strengthening (*38, 39*) and strain rate-induced hardening (*1*). These effects are likely active across all shock stresses, although they are overpowered by the softening mechanisms at sufficiently high stresses. Due to the difficulty of conducting control experiments at these conditions, we cannot directly distinguish the contributions of grain-boundary and strain rate effects. As shown in Fig. 2B, our measured flow strengths in NPD exceed those observed in shocked micro-polycrystalline diamond (*12*), though a quantitative determination of Hall-Petch behavior will require more precise measurements at other grain sizes.

Implications of stacking fault-mediated strengthening in diamond

Prior studies have achieved ultrahigh hardness in diamond under ambient conditions through grain boundary and nano-twin boundary strengthening (*14, 15, 31*). Our findings demonstrate a new stacking fault-mediated strengthening mechanism unlocked at multi-megabar pressures. Stacking faults, together with grain boundaries, reduce the mean free path of dislocations, thereby increasing the flow strength (*30*).

Notably, stacking fault-mediated strengthening was not observed in quasi-static compression experiments on NPD by Wang *et al.* (*7*). Although they measured a high ultimate flow strength of 23 GPa, their post-mortem microscopy revealed very few free dislocations in the recovered NPD samples. Instead, they found nano-twins, suggesting that at lower stresses, NPD deforms primarily via nano-twinning and grain boundary-mediated plasticity, with minimal dislocation activity within the grains. Recent experiments on single-crystalline diamond shocked to intermediate stresses of 69–115 GPa found both nano-twins and dislocations upon shock-recovery (*40*). Our MD simulations showed no clear evidence of nano-twin formation in the entire pressure range studied, suggesting that nano-twin formation may occur on longer timescales or upon pressure release. Our observation of dislocation-mediated plasticity in shocked NPD, in contrast to its absence under quasi-static compression, indicates that such mechanisms become accessible only at the much higher pressures achieved under shock loading.

The plasticity mechanisms we observed provide context for a new emerging understanding of pressure-dependent grain boundary strengthening in nano-grained materials. In nano-crystalline metals at ambient conditions, the deformation mechanism changes from dislocation slip to grain boundary sliding as grain size decreases, leading to a decline in strength known as inverse Hall-Petch behavior (*41*). Theoretical and experimental studies show that the application of pressure can prevent grain boundary sliding, inhibiting the Hall-Petch to inverse Hall-Petch transition (*30, 42, 43*). In nano-crystalline nickel, compressing to high pressure enables full dislocation slip to dominate the plasticity even down to 3 nm in grain size, leading to a maximum flow strength of 10.2 GPa via traditional grain boundary strengthening (*42*). MD studies of shocked nanocrystalline metals similarly show that high shock stresses can suppress grain boundary activity and promote full dislocation plasticity (*30, 43*), consistent with our shock results in NPD. However, our MD results at the highest shock stress of 623 GPa show substantial amorphization at the grain boundaries, suggesting that the high shock temperatures may enable grain boundary-mediated plasticity in addition to full dislocation motion.

Our findings extend to a broader class of ultrahard ceramic materials. While these materials are known to exhibit plasticity at quasi-static conditions of high pressure and temperature, few studies have been able to probe their plasticity mechanisms under extreme strain rates (*13, 44*). The measurement of grain reorientation to a <110> fiber texture provides evidence that diamond, the hardest among ceramics, can exhibit significant dislocation-based plasticity under shock compression. Moreover, our simulations demonstrate that the specific plasticity mechanisms evolve depending on the peak shock stress. These results pave the way for designing new ultrahard materials tailored to withstand specific deformation conditions for targeted applications including aerospace and nuclear energy.

## Materials and Methods

Experimental design

The nano-polycrystalline diamond (NPD) samples used in this work were synthesized at the Geodynamics Research Center, Ehime University. Highly oriented pyrolytic graphite was compressed to 15 GPa and heated to 2,300-2,500 ºC in the BOTCHAN 6,000-tonne multi-anvil press to form NPD.

The initial material density was 3.514 ± 0.003 g/cm$^3$, with an average grain size of 30-50 nm (*14, 15*). A 15-μm thick polypropylene film glued to the NPD sample was used as an ablator to produce a single shock front while minimizing emission of hard X-rays. The laser-driven shock compression experiments were performed using a high-intensity nanosecond drive laser at the BL3 beamline at SACLA (*16*). The laser was focused onto a 140 or 260 μm diameter spot on the target.

The lattice and microstructure of NPD were probed using horizontally polarized, ~10 fs X-ray pulses produced by the SACLA-XFEL. The X-rays were focused onto a 50 x 20 μm$^2$ spot on the target, which was tilted 108, 110, or 115 degrees with respect to the X-ray beam (Fig. 1A, Table S1). The X-ray energy was either 10, 11, or 12 keV, depending on the experimental run. X-ray diffraction (XRD) patterns were recorded on one or two pixel-array detectors. The delay times between the drive laser and XFEL probe pulses were tuned to probe the material within ~1 ns of the shock wave entering the NPD sample from the polypropylene ablator. This short time delay minimized stress gradients behind the shock wave at the probe time. Our experiment probed the entire sample depth. NPD's nano-grained microstructure (30-50 nm grains) allowed for clear and uniform diffraction measurements.

Strain under shock compression

Beyond the elastic limit of a material, the total strain $\varepsilon^t$ consists of two components: the elastic and plastic strains,

$$\varepsilon^t = \varepsilon^\ell + \varepsilon^p.$$

The elastic strain, which we call the lattice strain $\varepsilon^\ell$, is the direct deformation of the crystal lattice. The plastic strain $\varepsilon^p$ reflects irreversible deformation of the bulk.

Under shock compression, the total strain is uniaxial due to the directionality of the shock wave and lateral confinement from the surrounding material. This constrains the strain components,

$$\varepsilon^t_{\parallel} = \varepsilon^\ell_{\parallel} + \varepsilon^p_{\parallel},$$

$$\varepsilon^t_{\perp} = \varepsilon^\ell_{\perp} + \varepsilon^p_{\perp} = 0,$$

$$\varepsilon^\ell_{\perp} = -\varepsilon^p_{\perp},$$

$$\varepsilon^t_{\parallel} = \ln(\rho/\rho_0),$$

where $\varepsilon_\perp$ and $\varepsilon_\parallel$ are strain components along the shock direction and transverse to it, $\rho_0$ is the ambient density of NPD (3.514 g/cm$^3$) and $\rho$ is the density upon shock compression.

When the lattice deforms anisotropically, i.e., with greater compression along the shock direction than transverse to it, the projected strain on a given crystallographic plane depends on the orientation of that plane relative to the compression axis. As a result, diffraction from a given *hkl* reflection occurs at different d-spacings depending on the angle $\chi$ between the plane normal and the shock direction. These variations appear as warped diffraction patterns in polar plots of d-spacing versus azimuthal angle ($\phi$).

Behind the shock wavefront, plastic flow causes the material to yield and displace laterally. Because lateral confinement enforces a uniaxial total strain, the outward plastic strain is counteracted by a compressive lattice strain in the transverse directions ($\varepsilon^{\ell}_{\perp} = -\varepsilon^{p}_{\perp}$). This leads to a three-dimensional lattice strain. As the material loses its flow strength, it approaches the hydrodynamic limit where $\varepsilon^{\ell}_{\perp} = \varepsilon^{\ell}_{\parallel}$.

XRD Analysis

The first portion of XRD analysis was conducted following the method laid out in Hari *et al.* (*45*). Positions and orientations of the X-ray detectors were determined using diffraction from a $CeO_2$ calibrant and cross-checked with diffraction from the uncompressed NPD starting material. This calibration was first done in Dioptas (*46*), then refined using HEXRD (*47*). HEXRD was then used to transform the images to polar coordinates and perform Lorentz-polarization and pixel solid angle corrections.

The determination of lattice strains from the measured diffraction angle required multiple steps. The variation in 2θ with respect to the azimuthal angle was determined by dividing the de-warped diffraction patterns into 10-degree azimuthal slices. Each slice was azimuthally averaged, and the resulting one-dimensional lineout from each slice was fit to a PearsonVII peak shape. The lineouts of these slices are shown for each shot in Figs. S7–S15. The fit values of 2θ were then used to determine lattice strain and density of the shocked material, as described in the next section. The fits were also used to produce texture plots by observing the variation in the normalized area of the plastic peak with azimuthal angle. To ascertain the error on these fits, a final local least-squares refinement step is performed. From this least-squares step, the covariance matrix is calculated and used to extract fitting errors for each of the peak parameters. For further calculations, the errors were propagated in quadrature.

Determination of Lattice Strains and Flow Strengths

Diamond's strength leads to anisotropic lattice strain in the shocked sample, i.e., the lattice strain is greater along the longitudinal (shock) direction than along the transverse directions. We define $\varepsilon^{\ell}_{zz} = \varepsilon^{\ell}_{||}$ as the true longitudinal lattice strain component aligned with the direction of shock propagation, and $\varepsilon^{\ell}_{xx} = \varepsilon^{\ell}_{yy} = \varepsilon^{\ell}_{\perp}$ as the true transverse lattice strains perpendicular to the shock direction, which are assumed to be symmetric due to the uniaxial shock loading and the isotropic nature of NPD.

We determined $\varepsilon^{\ell}_{||}$ and $\varepsilon^{\ell}_{\perp}$ in the shocked NPD by using the variation in measured $2\theta$ values across different azimuthal slices. To do this, we performed an optimization in Matlab using the non-linear least squares method (lsqnonlin). The function that was optimized, Strain_Residuals.m, took in an initial guess for the lattice strain components and our experimentally measured $2\theta$ and $\phi$ values. By performing the calculation laid out below, this function used the guessed lattice strain components to calculate $2\theta$ values at each azimuthal angle. These calculated $2\theta$ values were compared to our experimentally measured $2\theta$ values, and the residuals between them were calculated, weighted by the uncertainties $\sigma_{2\theta}$ of the measured $2\theta$ values,

$$\text{residual}(\phi) = \frac{2\theta_{\text{model}}(\phi) - 2\theta_{\text{measured}}(\phi)}{\sigma_{2\theta_{\text{measured}}}(\phi)}.$$

These residuals were used as input to the lsqnonlin optimization method, which repeatedly ran Strain_Residuals.m with new inputs and optimized the lattice strain components to minimize the sum of squares of the residuals. Uncertainties in the lattice strain components were quantified by using a Monte Carlo approach of randomly choosing inputted values of measured 2θ from a normal distribution with a standard deviation equivalent to the measurement uncertainty, running the optimization 2000 times, and taking the mean and standard deviation of the outputted lattice strains.

The calculation of 2θ values from guessed lattice strain components is derived as follows. Due to the uniaxial planar shock loading, we take the shock and transverse directions as principal axes and assume no shear stresses or shear strains along these axes (*48, 49*). This approximation is justified due to our early probe timing (~1 ns after shock entry into the sample) while the shock is still planar. The deformation gradient tensor at the lattice level, $\mathbf{F}^{\ell}$, is thus diagonal in this coordinate system, and can be written as

$$\boldsymbol{F}^{\ell} = \begin{bmatrix} \exp(\epsilon^{\ell}{}_{\perp}) & 0 & 0 \\ 0 & \exp(\epsilon^{\ell}{}_{\perp}) & 0 \\ 0 & 0 & \exp(\epsilon^{\ell}{}_{\parallel}) \end{bmatrix}.$$

Our diffraction patterns contain traces that can be described by a diffraction angle 2θ and azimuthal angle ϕ. These angles together determine the normal direction of the diffracting plane. Let $\boldsymbol{n_0}$ be the unit vector normal to the diffracting plane in the undeformed (ambient) configuration. This initial normal direction is determined for each azimuthal sector from the 2θ of the undeformed crystal,

$$2\theta_0 = 2\arcsin\left(\frac{\lambda}{2d_0}\right).$$

and the ϕ value of that sector, from the relation

$$\hat{\mathbf{n}}_{\mathbf{0}} = \left(\sin\left(\frac{2\theta_0}{2}\right)\cos\phi\,, \sin\left(\frac{2\theta_0}{2}\right)\sin\phi\,, \cos\left(\frac{2\theta_0}{2}\right)\right).$$

Under deformation, the unit vector that describes the plane normal direction transforms via the inverse transpose of the deformation gradient,

$$\hat{\mathbf{n}}_{\text{strained}} = \frac{\mathbf{F}^{\ell^{-\mathbf{T}}}\hat{\mathbf{n}}_{\mathbf{0}}}{|\mathbf{F}^{\ell^{-\mathbf{T}}}\hat{\mathbf{n}}_{\mathbf{0}}|}.$$

Given the diagonal form of $\mathbf{F}^{\ell}$, this simplifies to

$$\hat{\mathbf{n}}_{\text{strained}} = \frac{\begin{bmatrix} \exp(-\epsilon^{\ell}{}_{\perp})n_x \\ \exp(-\epsilon^{\ell}{}_{\perp})n_y \\ \exp(-\epsilon^{\ell}{}_{\parallel})n_z \end{bmatrix}}{\sqrt{\exp(-2\epsilon^{\ell}{}_{\perp})(n_x^2+n_y^2)+\exp(-2\epsilon^{\ell}{}_{\parallel})n_z^2}}.$$

We define the angle $\chi$ as the angle between the strained plane normal and the shock direction, where

$$\cos\chi = \hat{\mathbf{n}}_{\text{strained}} \cdot \hat{\mathbf{z}}.$$

The lattice strain along the plane normal is then computed by projecting the full strain tensor onto $\hat{\mathbf{n}}_{\text{strained}}$, yielding

$$\epsilon_d^{\ell} = \epsilon_{\parallel}^{\ell} \cos^2\chi + \epsilon_{\perp}^{\ell} \sin^2\chi.$$

This leads to the change in d-spacing for the plane under the predicted strain,

$$d = d_0 \cdot \exp\left(-\epsilon_d^{\text{l}}\right),$$

where $d_0$ is the ambient d-spacing of the diffracting plane, and is given by

$$d_0 = \frac{a_0}{\sqrt{h^2+k^2+l^2}}.$$

In this equation, $a_0$ is the ambient lattice parameter of NPD (3.567 Å) and *h*, *k* and *l* are the miller indices of the diffracting plane.

Using Bragg's law, we obtain the predicted diffraction angle

$$2\theta = 2\arcsin\left(\frac{\lambda}{2d}\right),$$

where λ is the X-ray wavelength. This method produces results that are consistent with previous work that derived a closed-form equation relating the d-spacing and lattice strain components for arbitrarily strained samples (*48*).

Once the lattice strain components are optimized, density (ρ) of the strained sample is calculated from the ambient density ($\rho_0$) and lattice strain, as follows:

$$\rho = \frac{\rho_0}{\left(2-\exp\left(\varepsilon_{\parallel}^{\ell}\right)\right)\left(2-\exp\left(\varepsilon_{\perp}^{\ell}\right)\right)^2}.$$

We determined the flow strength $2\tau$ of the shock-compressed NPD using the relation

$$2\tau = \sigma_{\parallel} - \sigma_{\perp} = 2G\left(\varepsilon_{\parallel}^{\mathrm{l}} - \varepsilon_{\perp}^{\mathrm{l}}\right),$$

where G is the shear modulus. To estimate G, we used the Voigt iso-strain approximation from high-pressure elastic constants of diamond calculated with density functional theory (DFT) (*37*).

$$G = \frac{C_{11} - C_{12} + 3C_{44}}{5}.$$

Note that the flow strength is defined as twice the maximum shear stress $\tau$ (*50*). The flow strength is also equivalent to the differential stress sustained by the material during deformation.

Determination of Peak Shock Stresses

The peak longitudinal stress under plastic deformation ($\sigma_{\parallel}$) was determined for each run from the measured densities using the linear $U_S$-$u_p$ (shock velocity - particle velocity) relation of full-density NPD, which has been experimentally determined by Katagiri *et al.* (*51*). The samples used by Katagiri *et al.* were made by the same supplier as those used in the present work (*14, 51*).

The $U_S$-$u_p$ fit derived from the measurements was used along with the Rankine-Hugoniot conservation equations for mass ($\frac{\rho_0}{\rho} = \frac{U_s - u_p}{U_s}$) and momentum ($\sigma_{\parallel} = \rho_0 U_s u_p$) to determine longitudinal stress from density during shock propagation, $\sigma_{\parallel} = \rho u_p\left(U_s - u_p\right)$.

Forward Model Diffraction Simulations

Fiber diffraction patterns were simulated in Python using a proprietary GPU-accelerated Monte-Carlo method. Three-dimensional coordinates were defined with the convention of the z-axis pointing out of the page, x-axis pointing to the right, and y axis pointing up. All calculations were done from a reference frame in which the beam direction was defined as pointing along the -z axis. The shock direction was defined as a 108-, 110- or 115-degree rotation of the beam direction about the x axis (Fig. 1A, Table S1).

First, diamond's crystallographic planes were defined as unit vectors pointing in the direction normal to the plane. These plane normal vectors were then rotated such that the crystallographic direction corresponding to the fiber axis was aligned with the shock direction. The Monte-Carlo engine assigns a random (uniformly distributed) twist between 0° and 360° and a mosaic spread away from the fiber axis (using a gaussian distribution with a standard deviation of 10°) to each grain. For every grain, the X-ray beam was analytically reflected from each plane; if the reflection's wavelength satisfied

$$|\lambda - 2d \sin \theta_b| < \Delta\lambda,$$

where $\Delta\lambda$ corresponds to the X-ray bandwidth. Corresponding polar (2θ) and azimuthal (φ) angles are then recorded, provided they also lay within the detector window. In this case $5 \times 10^9$ grains were simulated. Scherrer broadening of these recorded points for 30 nm crystallites is then applied. It must be noted that we are not claiming that the NPD reached an orientation distribution where the greatest misorientation angle from the fiber axis was 10°. This degree of mosaicity was chosen for ease of comparison between the data and simulations. Since the NPD began with a fully random orientation distribution, it is to be expected that some portion of the texture remains random, even as a preferred orientation emerges.

## Large-Scale Molecular Dynamics Simulations

A series of large-scale MD simulations were carried out to investigate the shock response of nanocrystalline Diamond using LAMMPS (*52*). A quantum-accurate machine learning, Spectral Neighbor Analysis Potential (SNAP)-ML-IAP for diamond (*53*) was used in our work. This machine learning potential has been demonstrated in its capability of describing diamond's properties at extreme conditions spanning pressures from 0 to 5,000 GPa and temperatures up to 20,000 K, including the phase diagram and melting curve of diamond with high accuracy. We also calculate the Hugoniot curve of NPD, including the pressure versus particle velocity, which shows good agreement with experimental measurements from Katagiri *et al.* (*51*) (see Fig. S2).

The simulated NPD samples are generated using a Voronoi-construction method. In order to eliminate any size effect resulting from the number of grain boundaries in the cross section of the sample and to ensure sufficient number of grains for texture analysis, the length of the simulated system is ~50.5 × 50.5 × 150.8 $nm^3$ along x-, y-, and z- dimensions, respectively, with the number of atoms up to 66 million. The initial model contains about 900 grains with an average grain size of ~ 8 nm. Prior to shock loading, the simulated systems are relaxed under the isothermal-isobaric (NPT) ensemble which maintains constant temperature and constant pressure applied, and equilibrated at 300 K for 30 ps.

In shock MD simulations, we adopt a "piston" method which uses a virtual piston wall that impinges upon the sample such that the particle velocity in the sample is the same as the piston speed after the shock reaches steady state. In detail, with periodic boundaries along the x, and y dimensions and a free surface along the z direction, planar shock loading is achieved along the z direction by introducing a flat-surface, infinite-mass piston moving in the positive z direction from the lower z boundary. During the whole shock loading, the system is maintained under a microcanonical (NVE) ensemble where the particle number (N), volume (V), and energy (E) are consistent. corresponding to an isolated system that cannot exchange heat or matter with the outer environment. A timestep of 1 fs is used. This methodology creates shock with uniaxial total strain. The shock wave is held as it propagates to the rear free surface along the z direction.

The virial stress definition is used, where the stress tensor component for atom $i$ is

$$\sigma_{\alpha\beta} = -\frac{1}{\Omega_i}[m_i v_{i,\alpha} v_{i,\beta} + \frac{1}{2}\sum_{n=1}^{N_p}(r_{1\alpha}F_{1\beta} + r_{2\alpha}F_{2\beta}) + \frac{1}{3}\sum_{1}^{N_a} r_{1\alpha}F_{1\beta} + r_{2\alpha}F_{2\beta} + r_{3\alpha}F_{3\beta}],$$

where $\Omega_i$ , $m_i$ and $v_i$ are the atomic volume, mass, and velocity of atom $i$, respectively. The first term is a thermal kinetic energy contribution, in which the binned center-of-mass translational velocity has been subtracted. In the second term, $n$ loops over the $N_p$ pairwise neighbors of atom $i$, and the third term is for the $N_a$ angular (three-body) interactions that involve atom $i$. In these terms, $r_1$, $r_2$, and $r_3$ are the positions of atoms in the pairwise or three-body interactions, and $F_1$, $F_2$, and $F_3$ are the resulting pairwise or three-body forces on those atoms.

By averaging among a group of atoms and subtracting the local center-of-mass velocity along the shock direction, we calculate the temperature

$$T = \frac{1}{2Nk_b}\sum_{i=1}^{N} m_i(v_{i,x}^2 + v_{i,y}^2),$$

where $k_b$ is Boltzmann's constant, $m_i$ is the atomic mass of atom $i$, and $v_{i,x}$ ; $v_{i,y}$ are components of the particle velocity.

Under uniaxial strain compression, the longitudinal stress ($\sigma_{\|}$) along the loading axis (Z) is the shock stress, and there are also non-zero stress components along the lateral directions, denoted as the X and Y axes, respectively. The hydrodynamic pressure is calculated as

$$P = -\frac{\sigma_{11} + \sigma_{22} + \sigma_{33}}{3}$$

and the von Mises flow stress is calculated using

$$\sigma_{\text{flow}} = \sqrt{\frac{(\sigma_{11}-\sigma_{22})^2+(\sigma_{22}-\sigma_{33})^2+(\sigma_{33}-\sigma_{11})^2}{2}}.$$

Notably, when $\sigma_{11} = \sigma_{22}$, which holds for isotropic materials under uniaxial strain, the flow stress simplifies to

$$\sigma_{\text{flow}} = \sigma_{33} - \sigma_{11} = \sigma_{\parallel} - \sigma_{\perp}.$$

This is analogous to the expression used to determine flow stress from the experiments.

To analyze the shock response in the MD simulations, the system is divided into bins along the z-axis shock direction in which local physical quantities are averaged, including components of the stress tensor, particle velocity, temperature, density, etc. Each bin has a width of twice the lattice constant (~3.57 Å at 300 K). OVITO (*54*) is used in all visualizations in this work. Many techniques including Identify diamond structure, Dislocation analysis (DXA), Polyhedral template matching, Grain segmentation are utilized in microstructure analysis of NPD. Inverse pole figures and statistics of misorientation angle to <110> are used for texture analysis.

Diffraction Patterns Simulated from Molecular Dynamics Simulations

The atomic position file outputs from the MD simulations were used as inputs to a proprietary GPU-accelerated Python atomistic wave-optics simulation. To gain better statistics, a number of additional MD simulation outputs were combined to form a total sample with 1180 grains.

This simulation initializes an X-ray beam as a finite ~$10^{-3}$ bandwidth plane wave and tracks the individual scattering of the wave on each atom based on

$$E_s = E_i\big(f_0(|\vec{q}|) + f_1(\lambda) + if_2(\lambda)\big),$$

where $f_0, f_1, f_2$ are the atomic scattering factors of each species which alter the magnitude and phase of the wavefield. Scattered wavefields are then propagated and accumulated on the detector pixels.

The imported atomic positions were aligned so that the shock axis and incoming beam direction matched the experimental conditions.

**Acknowledgments**

We thank Paul E. Specht of Sandia National Laboratories, Patrick G. Heighway of the University of Oxford, and Darshan Chalise, Dayeeta Pal and Vittal Bhat of Stanford University for scientific discussions. The experiments were performed at BL3 of SACLA with the approval of the Japan Synchrotron Radiation Research Institute (proposal nos. 2024A8007, 2023A8016 and 2019A8041). The NPD sample fabrication was conducted under the support of Joint Research Center PRIUS (Ehime University, Japan). The high-power drive laser installed in SACLA EH5 was developed with the corporation of Hamamatsu Photonics. The installation of a Diffractive Optical Elements (DOE) to improve the smoothness of the drive laser pattern was supported by the SACLA Basic Development Program. The MD simulations were performed at A*STAR Computational Resource Centre (A*CRC).

**Funding**

Air Force Office of Scientific Research Contract FA9550-23-1-0347

MEXT Quantum Leap Flagship Program (MEXT Q-LEAP) Grant No. JPMXS0118067246

Japan Society for the Promotion of Science (JSPS) KAKENHI (Grants Nos. 21J10604, 19K21866, & 16H02246)

Genesis Research Institute, Inc. (Konpon-ken, Toyota)

Yamada Science Foundation

**Author contributions**

Conceptualization: KK, NO

Sample preparation: TI, TS, NN

Experimentation: KK, AH, TR, NO, LEDM, AA, YI, SJI, MK, LM, TM, KM, HN, NN, TO, MO, TS, YS, KS, YT, ST, TT, YU, MY, TY

Experimental data analysis: AH, DL, RL

Data analysis software: AH, RH, DL

MD simulations: WL

Diffraction simulations: DL, AH

Writing: AH, WL, DL, SEP, ZX, KK, LEDM

Figures: AH, WL, RL, DL, EWC

Supervision: LEDM, KK, NO, RK

**Data and materials availability**

All data are available at

https://drive.google.com/drive/folders/1KcgtJ68ZcWDmqkmf1D8ZjpSg6tZnpy_T?usp=sharing.

## Competing interests

The authors declare that they have no competing interests.

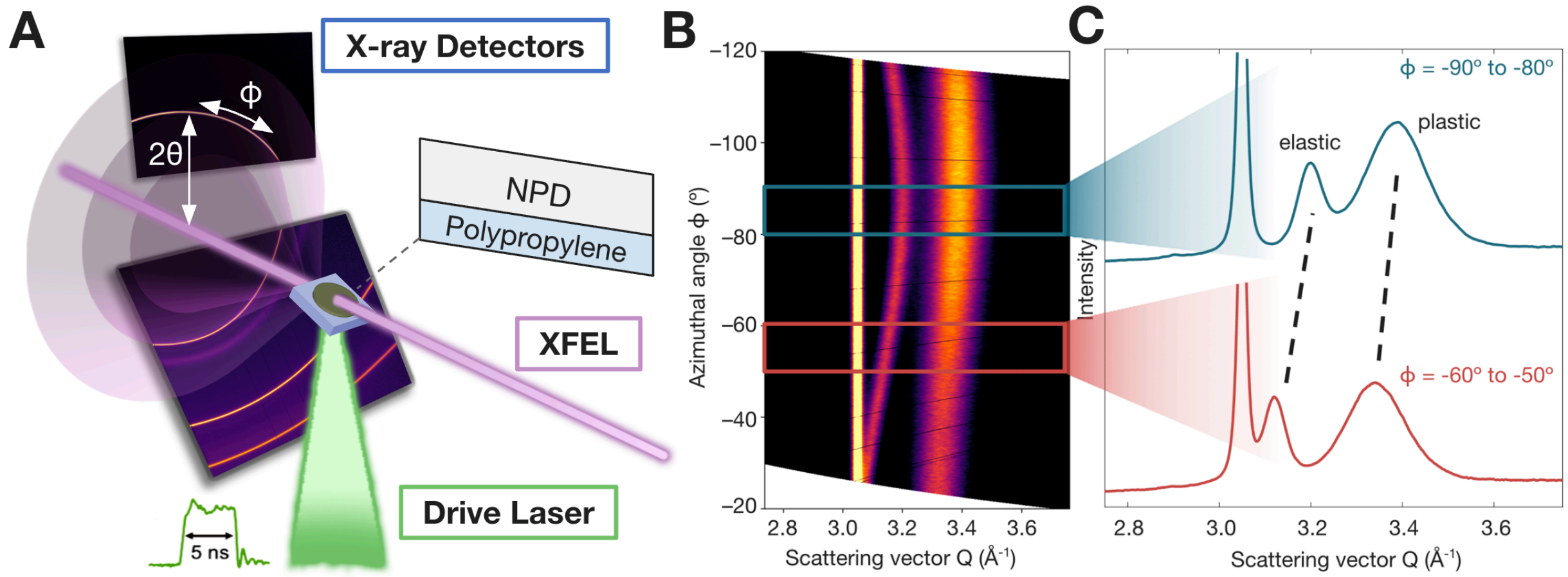

**Fig. 1. Shock experiments with femtosecond X-ray diffraction.** (**A**) Laser driven shock experiments coupled with *in situ* X-ray diffraction (XRD) using a femtosecond XFEL pulse irradiating the sample at an angle relative to the shock propagation direction. (**B**) Representative XRD pattern of shocked NPD. The XFEL probes the uncompressed, elastically compressed and plastically compressed states at once. Higher Q values correspond to greater degrees of compression. (**C**) 1D profiles are shown from specific azimuthal slices.

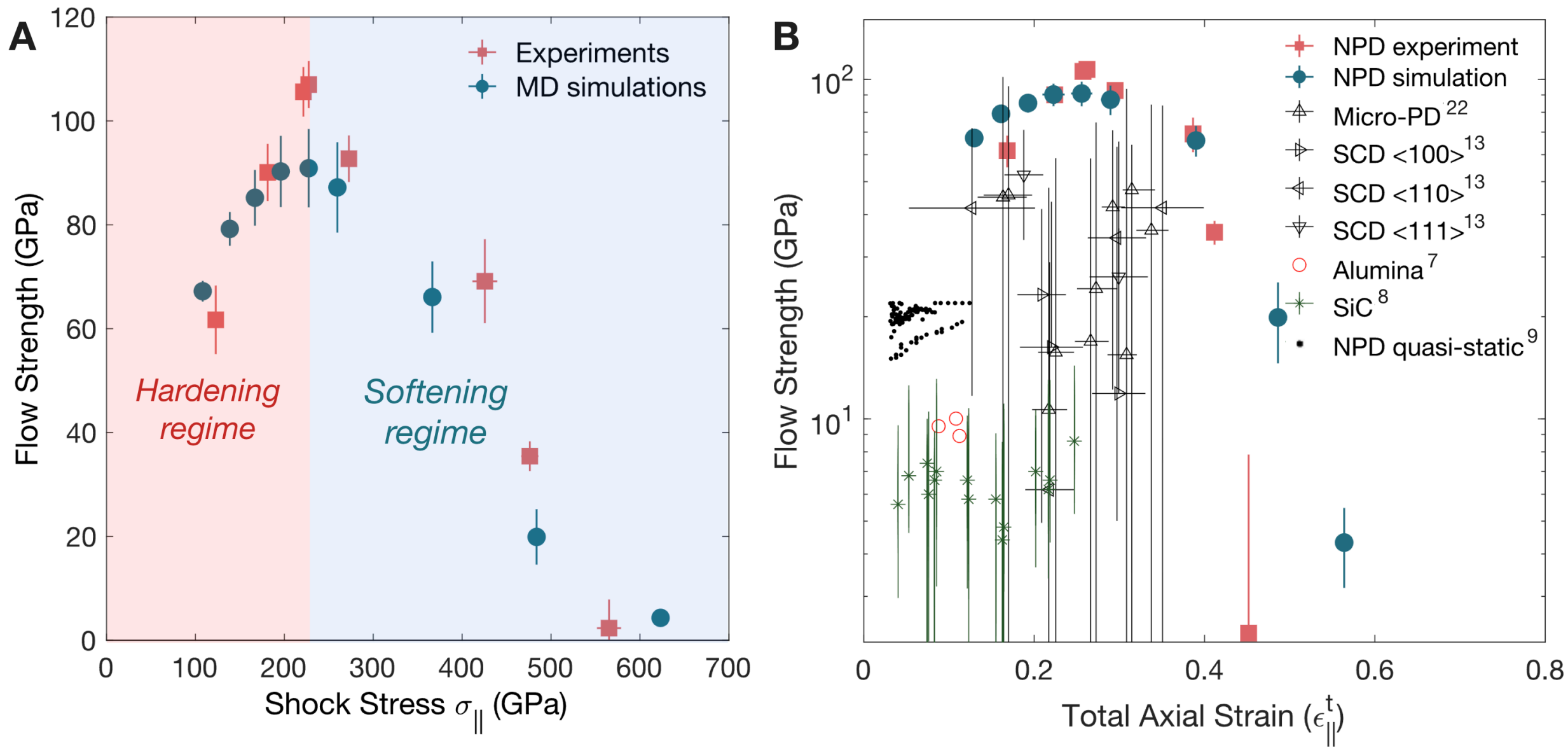

**Fig. 2. Ultra-high flow strength of shocked NPD.** (**A**) Flow strength as a function of peak shock stress. Both experiments and simulations show stress-hardening up to 227 ± 8 GPa and stress-softening at higher stresses. (**B**) Logarithmic plot comparing shocked NPD's flow strength to that of other shocked materials as well as quasi-statically compressed NPD. Our measured flow strength is higher than previous measurements under shock loading and represents a five-fold increase compared to that of quasi-statically compressed NPD.

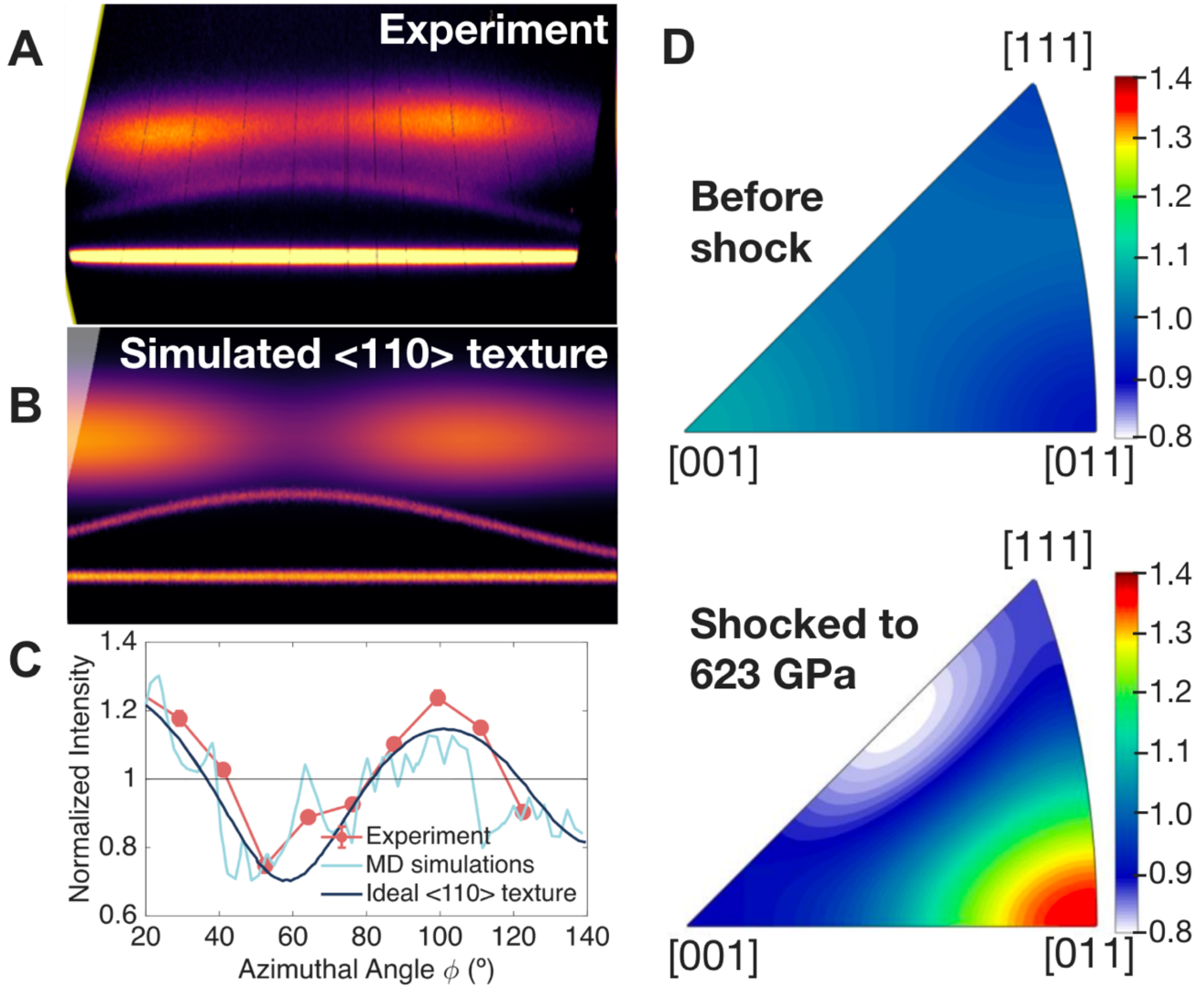

**Fig. 3. Ultrafast texture formation.** (**A**) In the high stress softening regime, diffraction from the shocked material displays azimuthal variation in intensity that is not observed in the uncompressed material. This indicates grain reorientation, i.e., formation of crystallographic texture. (**B**) Forward model of a <110> fiber texture with a mosaic spread, where the <110> directions align with the shock axis and directions normal to it are randomly distributed. (**C**) Line plots comparing the forward model of <110> texture with the experimental and MD data as integrated intensity along ϕ. The data show good agreement with the model of <110> fiber texture. (**D**) Inverse pole figures

generated from molecular dynamics simulations at 623 GPa also indicate the formation of <110> fiber texture.

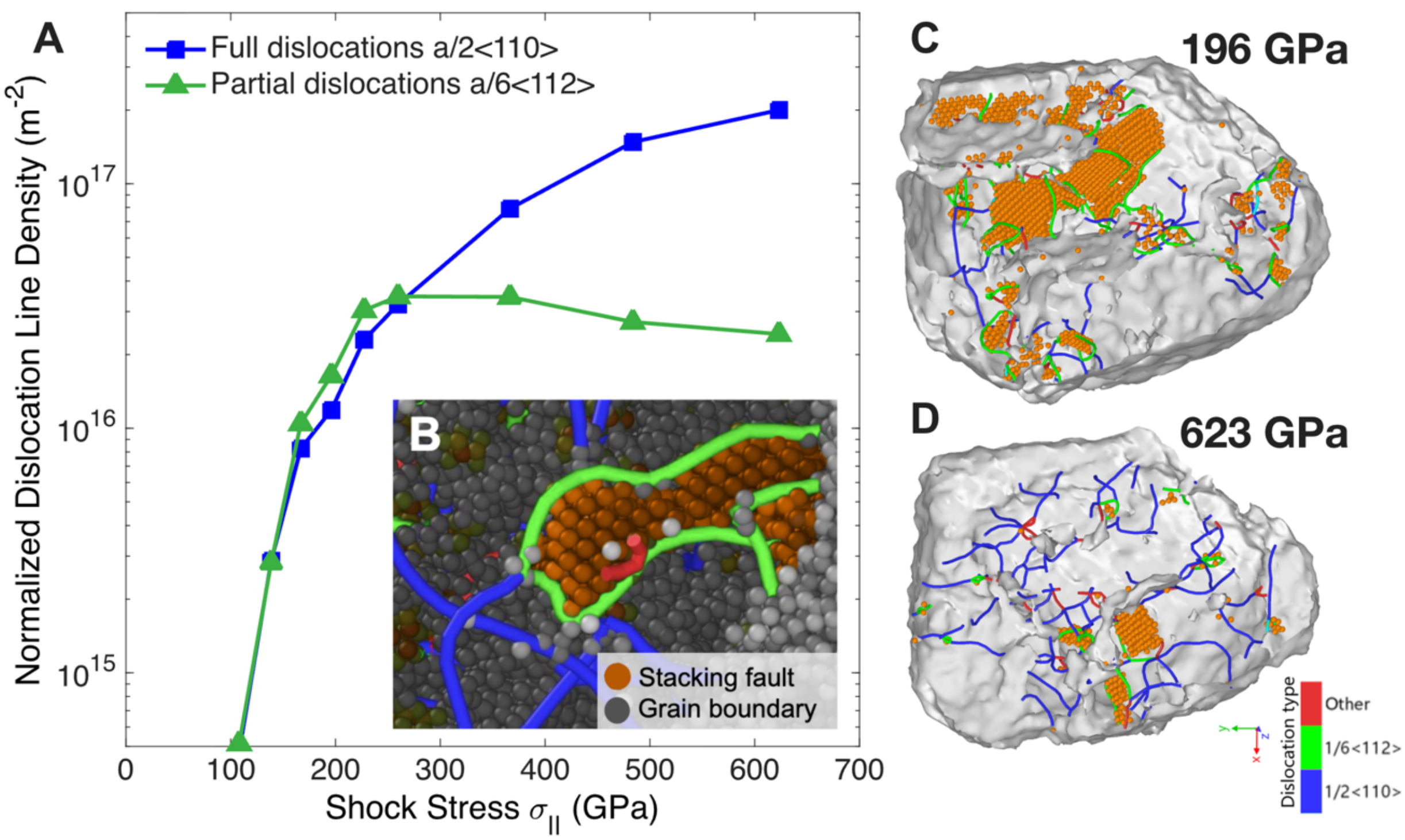

**Fig. 4. Evolution of dislocation character in molecular dynamics simulations.** (**A**) Densities of a/6<211> partial and a/2<110> full dislocations with increasing shock stress. The dislocation line densities are normalized to the proportion of cubic diamond atoms in the system. Up to 260 GPa, partial and full dislocation densities increase with shock stress, with a higher density of partials. At higher stresses, the full dislocation density continues to increase, while the partial dislocation density slightly reduces. (**B**) Snapshot of a full dislocation (blue) dissociating into two partial dislocations (green) separated by a stacking fault (orange atoms). (**C**) Snapshot of a grain shocked to 196 GPa shows partial dislocations, stacking faults, and a lower proportion of full dislocations. (**D**) Snapshot of the same grain shocked to 623 GPa shows a much higher density of full dislocations and fewer stacking faults. For ease of visualization, cubic diamond atoms are not shown in **B**, **C** and **D**.

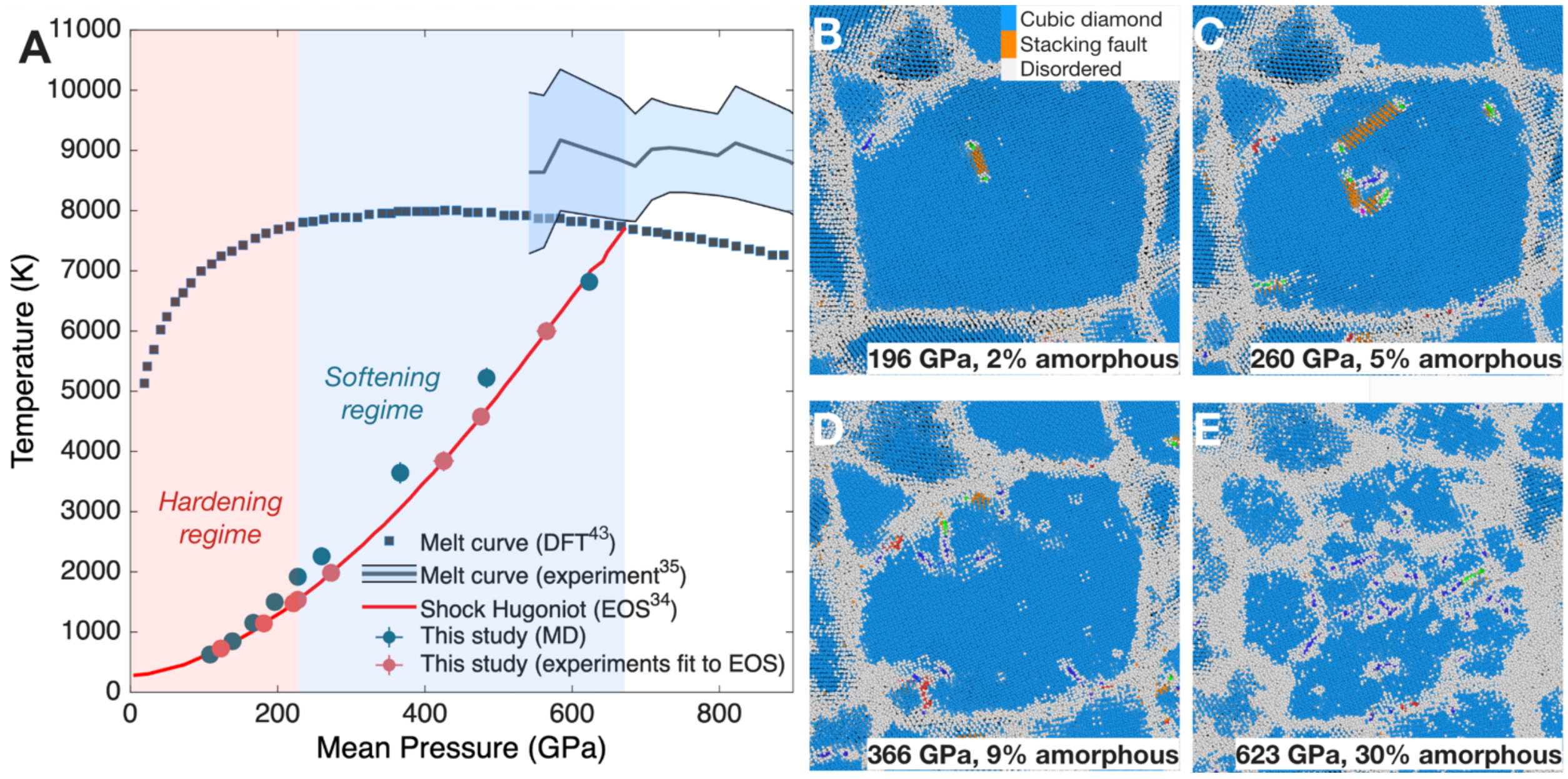

**Fig. 5. Shock temperatures and grain boundary thickening.** (**A**) Shock temperatures were determined for the experimental data by fitting to a high-pressure carbon equation of state (EOS) (*34*). Temperatures were calculated from MD simulations using the two atomic velocity components perpendicular to the shock direction. Melt curves from shock experiments (*35*) and DFT simulations (*43*) are shown. The fit temperatures in the high-stress softening regime reach as high as 6000 ± 150 K at 565 ± 14 GPa, approaching the melt. (**B–E**) Grain-level snapshots from MD simulations are shown at (**B**) 196 GPa, (**C**) 260 GPa, (**D**) 366 GPa and (**E**) 623 GPa. Full dislocations are shown in blue and partials are shown in green. Increasing proportion of disordered atoms (gray) causes grain boundary thickening at higher stresses.

## Supplementary Materials

<u>Differences between experimental and MD results</u>

Although our experimental and MD results show good agreement overall, there are a few differences in the results. The experiments show a sharper rise and fall in flow strength, and the NPD loses nearly all strength at a shock stress of 565 ± 8 GPa. By contrast, the simulations maintain significant strength of 20 ± 5 GPa even up to a shock stress of 484 GPa. Furthermore, the onset of texture formation in the experiments occurs at 425 ± 14 GPa, while in the MD simulations it occurs only at 623 GPa. We attribute these slight discrepancies to the differences in grain size and timescale between experiments and simulations. The MD simulations used a grain size of 8 nm and measured the shock response over several picoseconds, while experiments used 30-50 nm grains and measured ~1 ns after shock entry into the NPD. The smaller grain sizes of MD simulations may yield a slightly lower strength because the fraction of grain boundaries increases significantly compared to grain interiors, thereby lowering the shear modulus of the material (*55*).

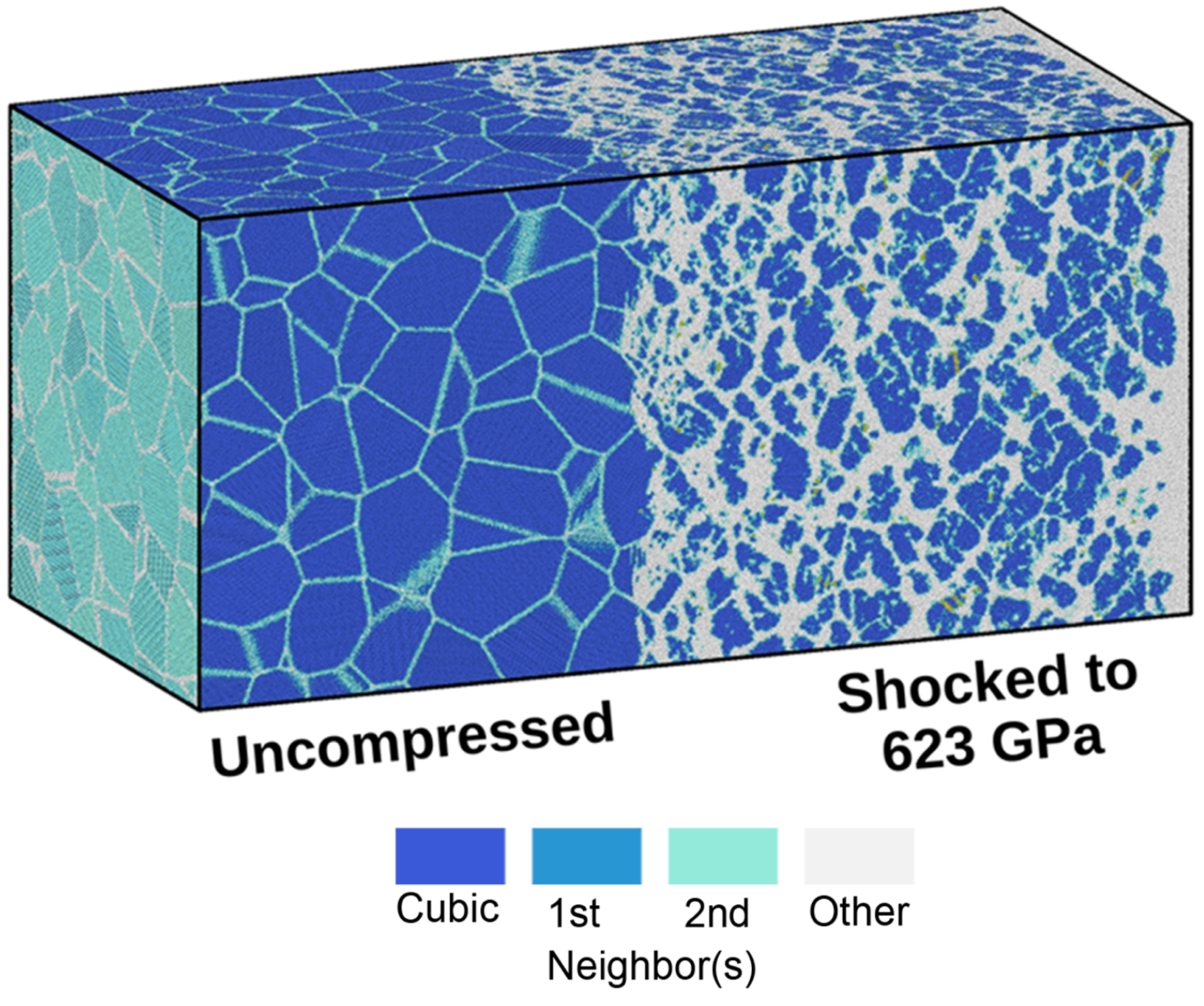

**Fig. S1. Large-scale molecular dynamics simulations.** Molecular dynamics simulation consisting of 900 diamond grains shocked to 623 GPa.

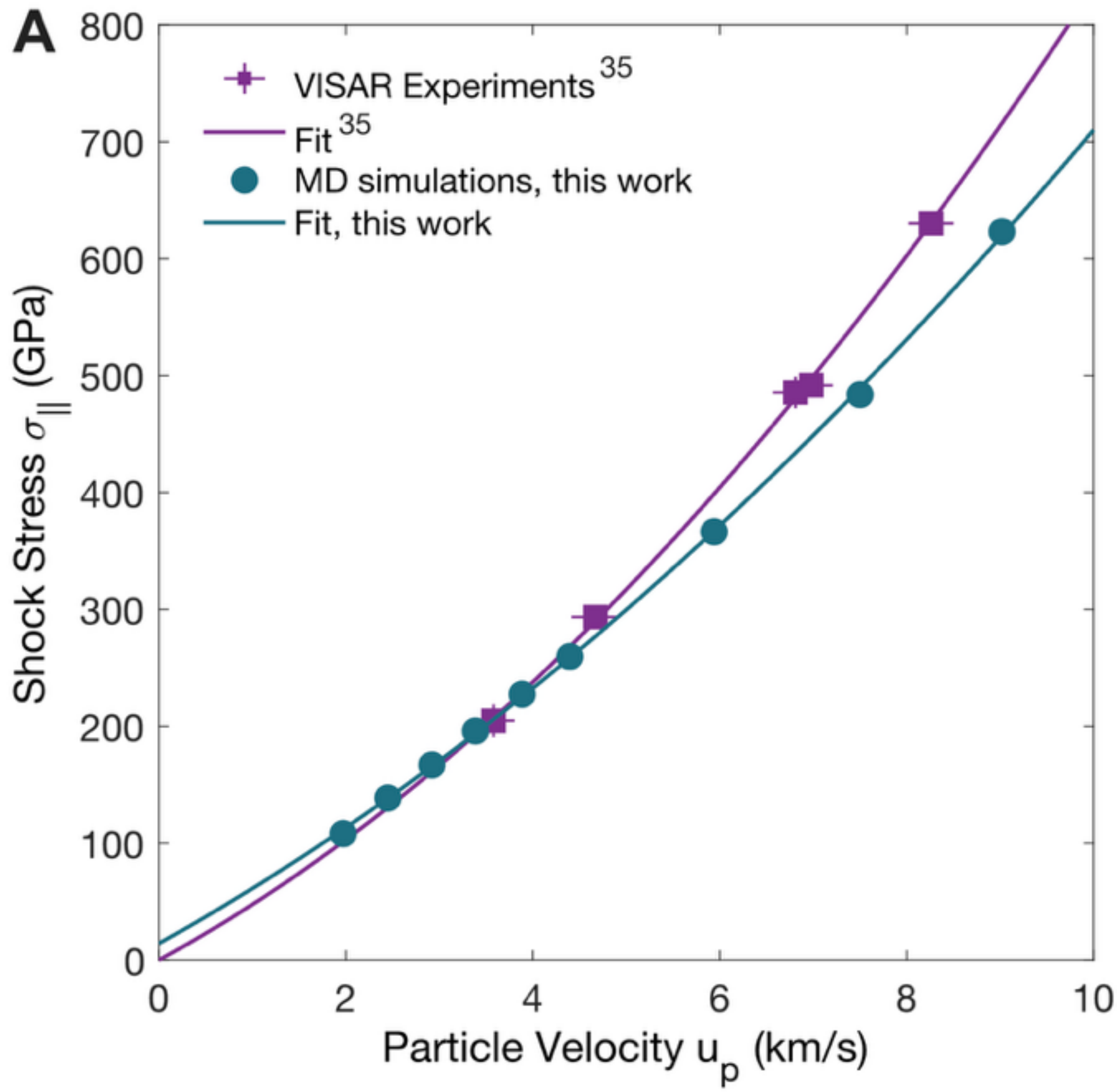

**Fig. S2. Shock Hugoniot data from MD simulations.** (**A**) Shock Hugoniot curve of NPD from MD results of this work, compared with experimental Hugoniot results on NPD (*51*). Shock stress is shown as a function of particle velocity. We determined the shock stress in our experiments using the Hugoniot results from (*51*).

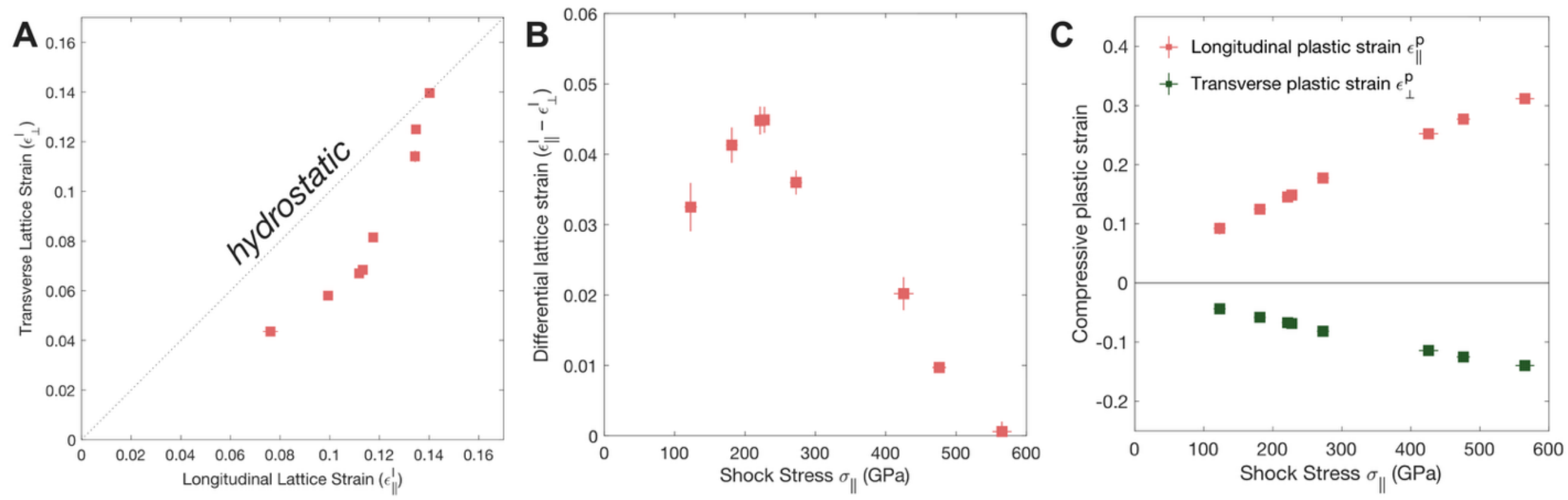

**Fig. S3. Lattice strains from *in situ* XRD.** (**A**) Longitudinal vs transverse lattice strains are shown for our experimental results and compared to hydrostatic compression (zero strength). (**B**) Differential lattice strains from our experiments are plotted as a function of peak stress. (**C**) Longitudinal and transverse plastic strains, calculated from the measured lattice strains, are plotted as a function of peak stress. All strains shown are compressive strains, i.e., positive values indicate compression while negative values indicate expansion.

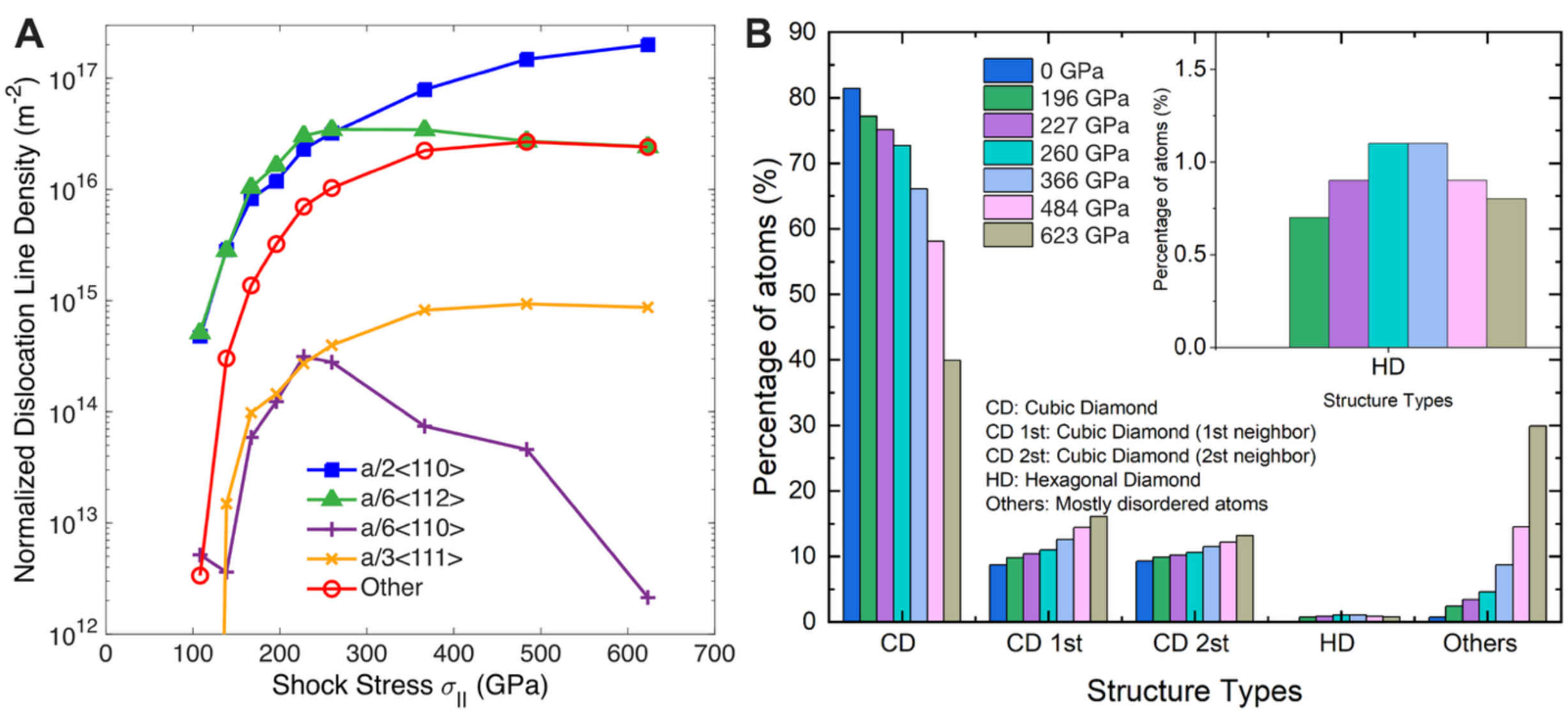

**Fig. S4. Crystal structure and defects.** (**A**) Dislocation densities from MD simulations as a function of shock stress for various dislocation characters. The densities are normalized to the proportion of atoms in the cubic diamond structure (including first and second neighbors) at each shock stress. (**B**) Proportion of atoms in different structures. The hexagonal diamond structure results from stacking faults in the material.

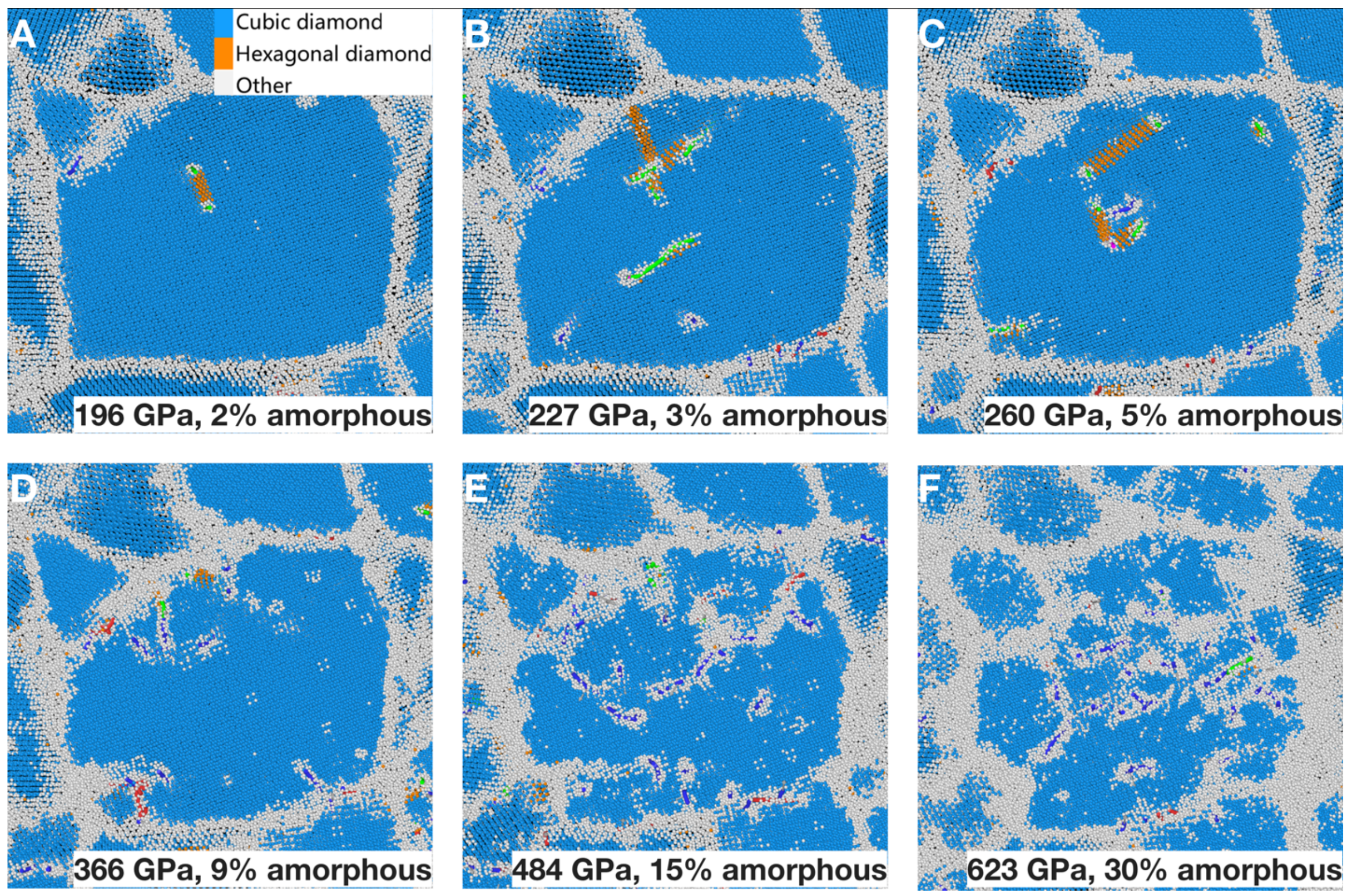

**Fig. S5. Grain boundary thickening.** Grain-level snapshots from MD simulations at (**A**) 196 GPa, (**B**) 227 GPa, (**C**) 260 GPa, (**D**) 366 GPa, (**E**) 484 GPa and (**F**) 623 GPa.

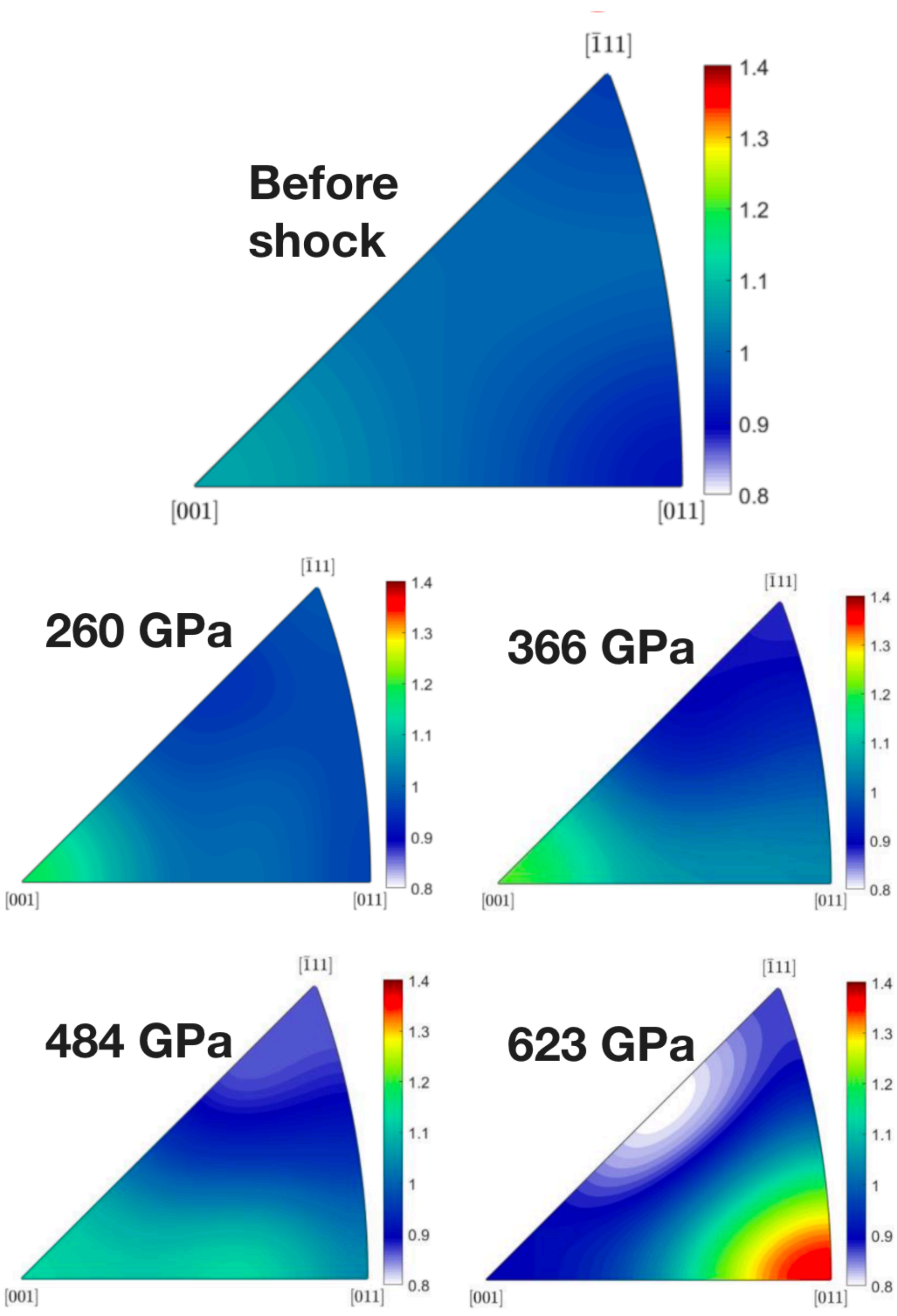

**Fig. S6. Grain reorientation from MD data.** Inverse pole figures from MD data at various shock stresses.

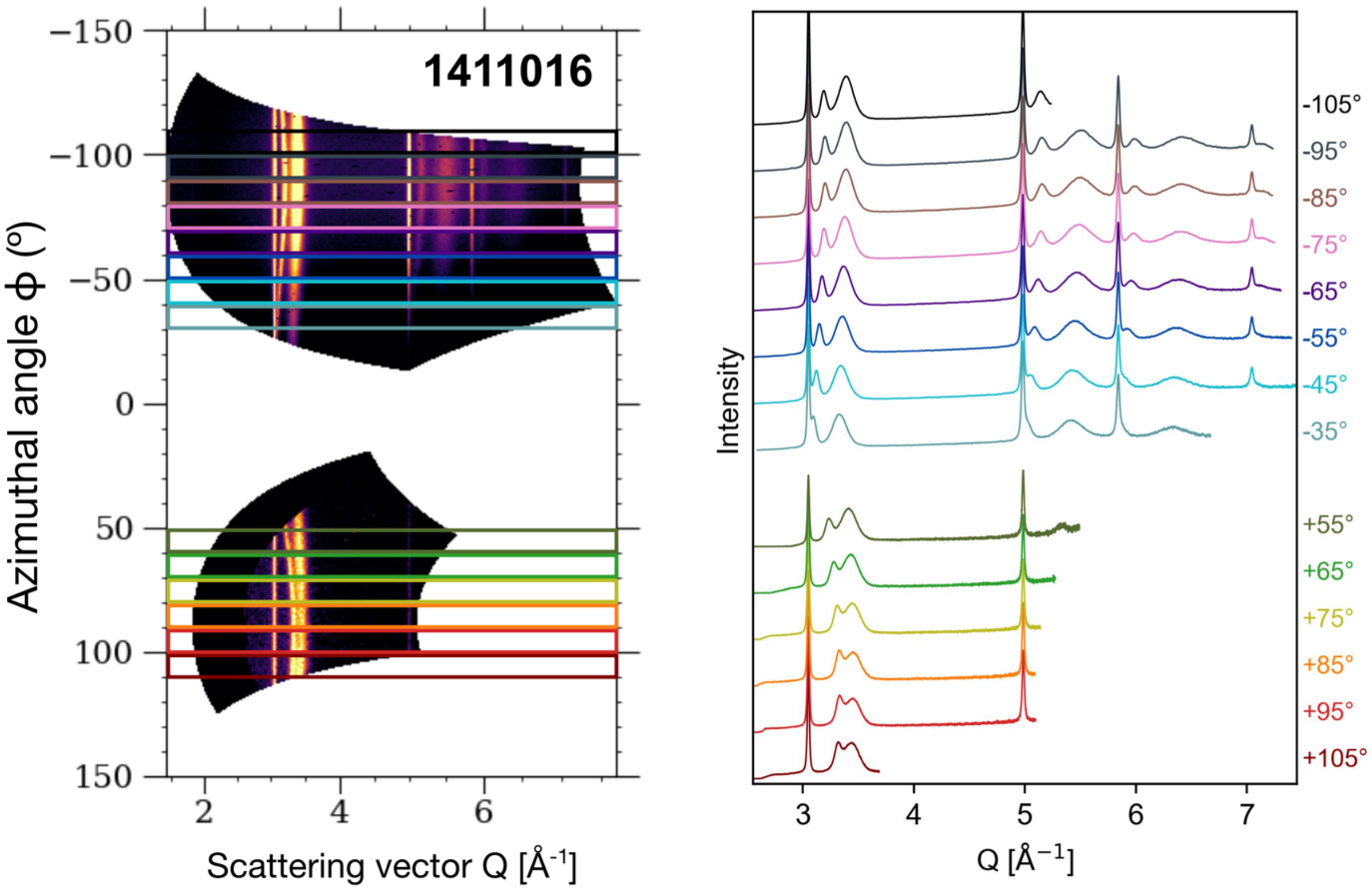

**Fig. S7. Example diffraction pattern and azimuthal slices.** Shown is an example diffraction pattern of NPD shocked to 227 ± 8 GPa, with flow strength of 107 ± 5 GPa. To determine the differential lattice strains used to calculate strength, the diffraction pattern was sliced into 10-degree azimuthal sectors, shown to the right. The flow strength is determined from the degree of variation of the scattering vector Q with respect to azimuthal angle.

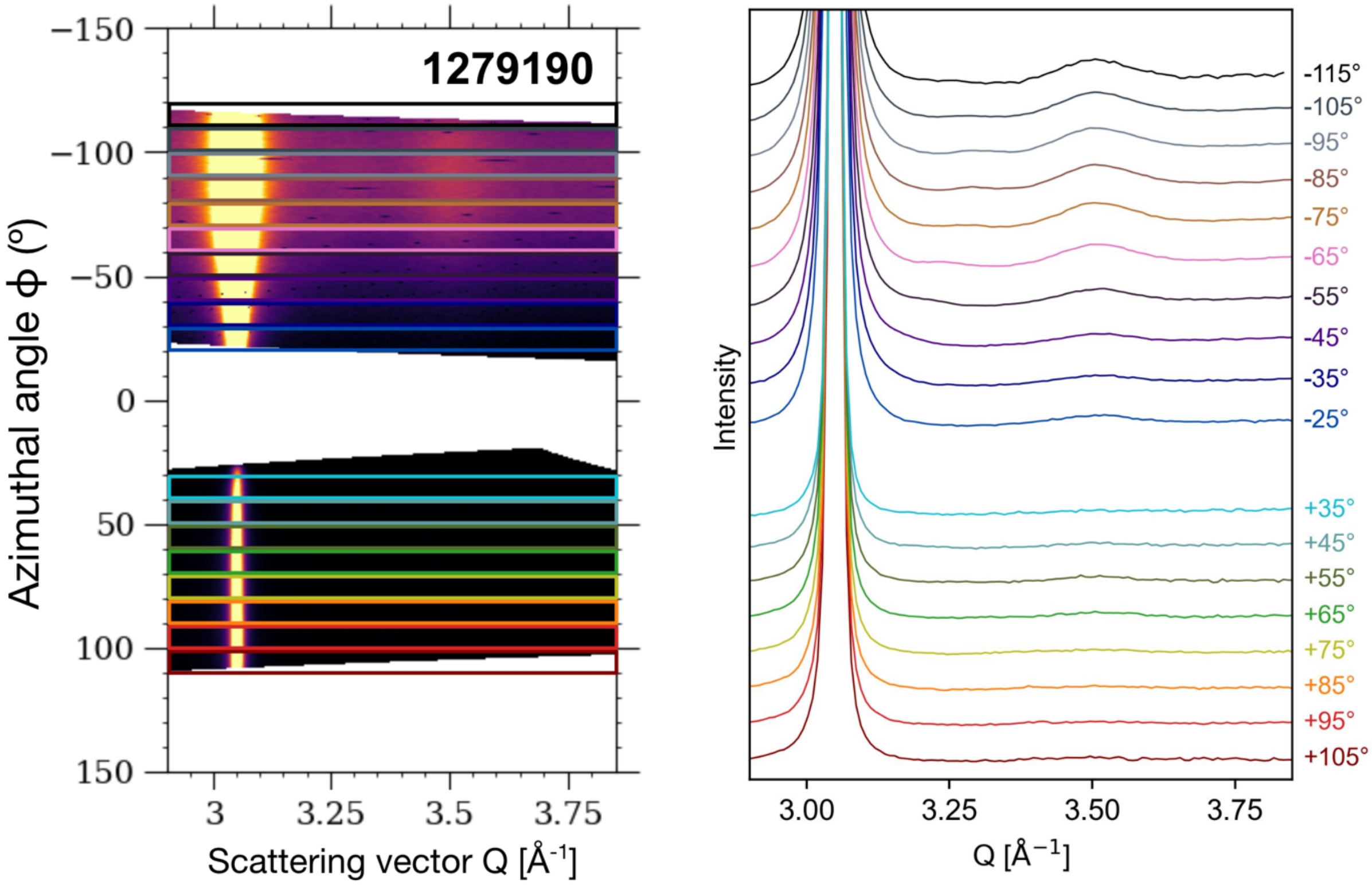

**Fig. S8.** Diffraction pattern of Run 1279190 (NPD shocked to 565 ± 14 GPa), with lineouts from each 10-degree azimuthal sector. To show the azimuthal variations more clearly, only diffraction from the {111} planes is included.

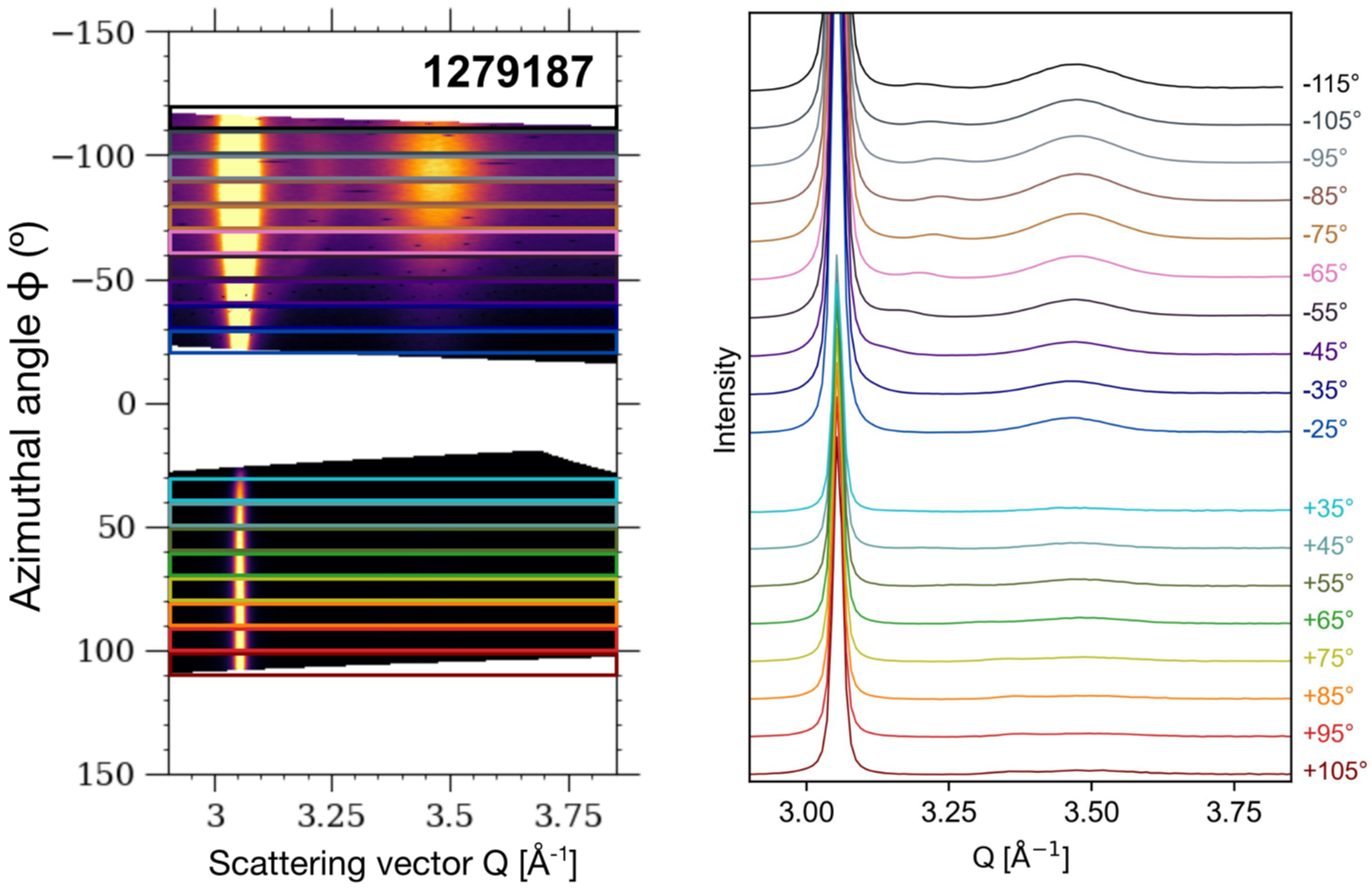

**Fig. S9.** Diffraction pattern of Run 1279187 (NPD shocked to 476 ± 10 GPa), with lineouts from each 10-degree azimuthal sector. To show the azimuthal variations more clearly, only diffraction from the {111} planes is included.

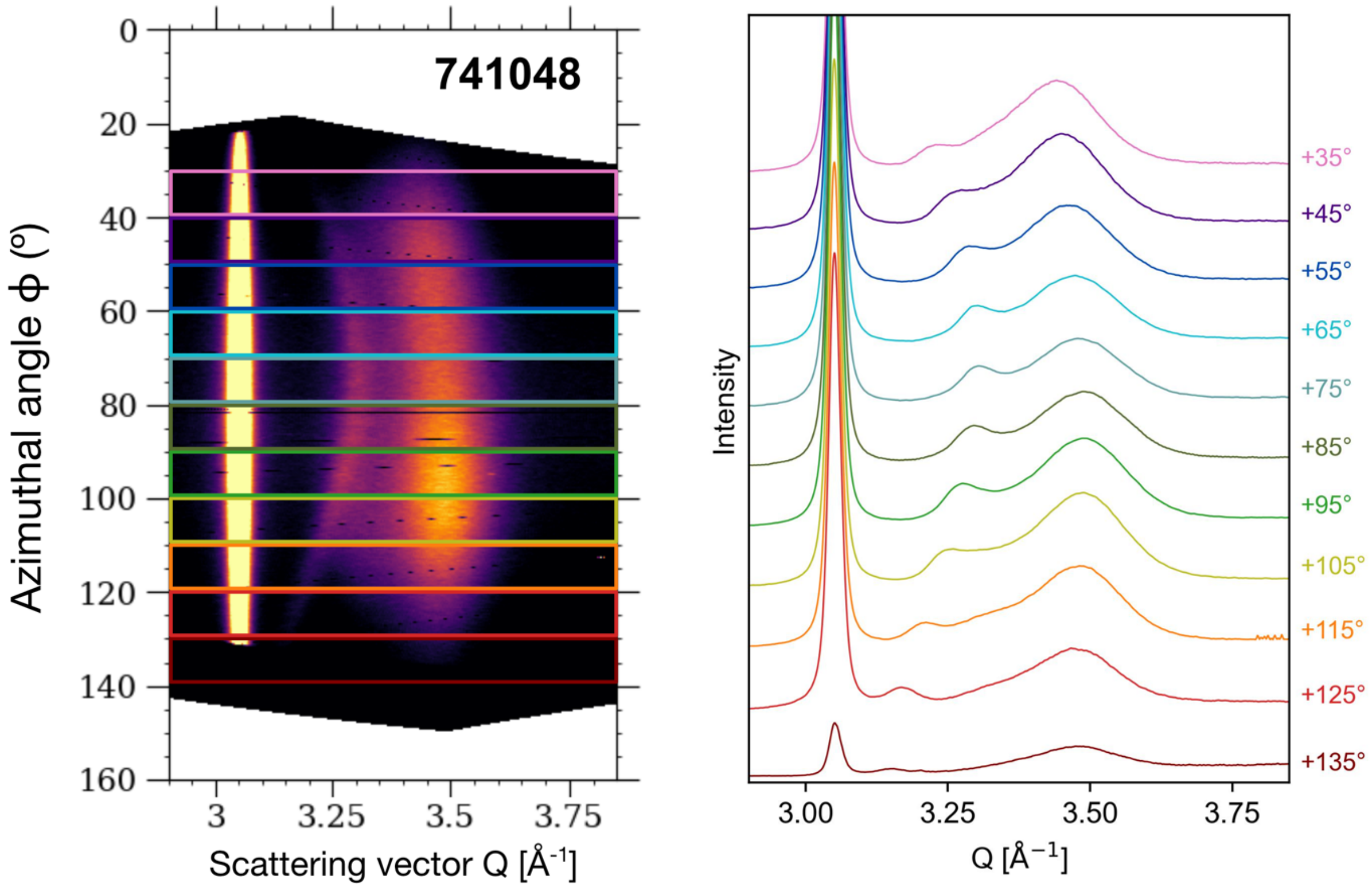

**Fig. S10.** Diffraction pattern of Run 741048 (NPD shocked to 425 ± 14 GPa), with lineouts from each 10-degree azimuthal sector. To show the azimuthal variations more clearly, only diffraction from the {111} planes is included.

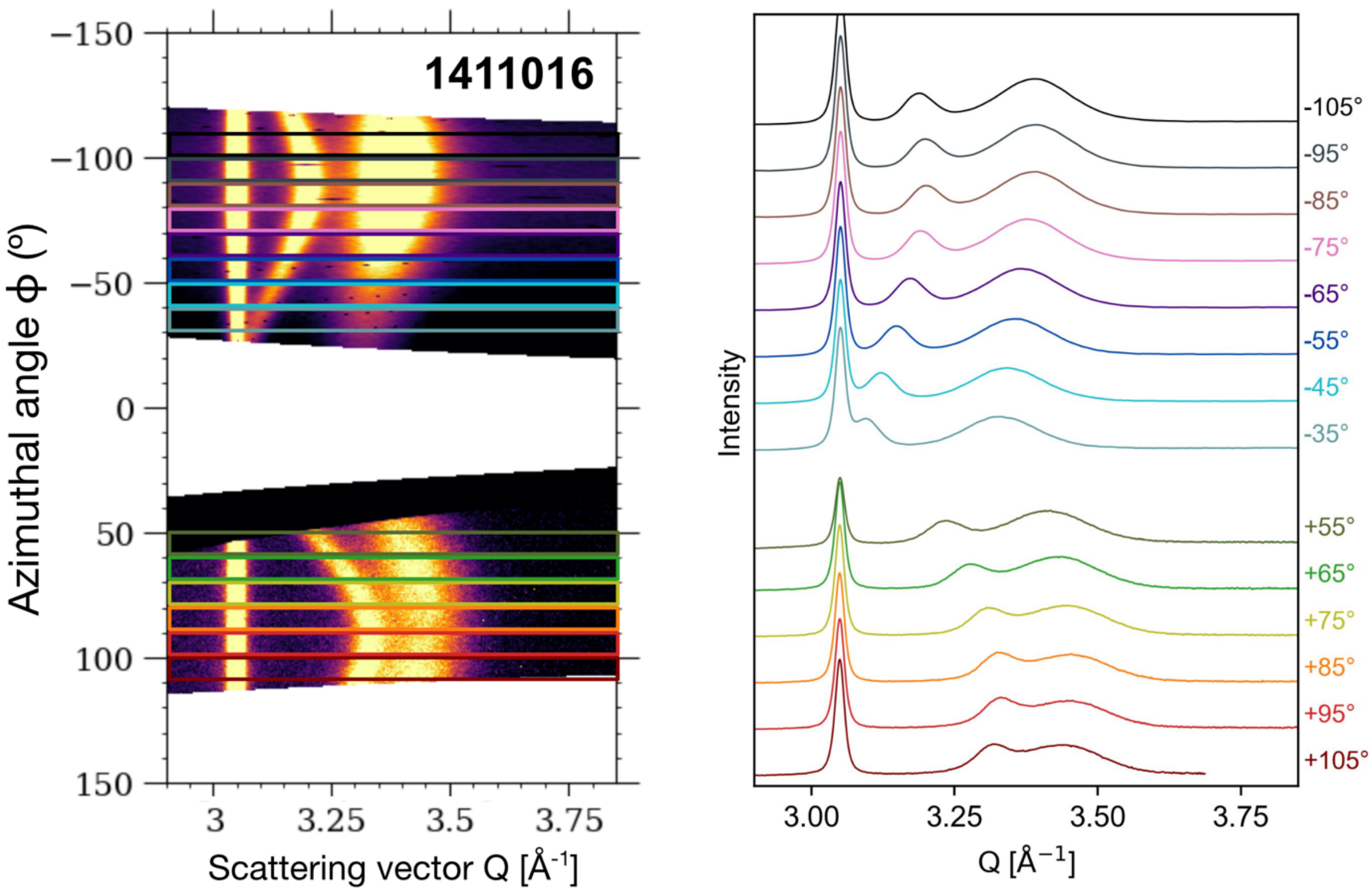

**Fig. S11.** Diffraction pattern of Run 1411016 (NPD shocked to 273 ± 9 GPa), with lineouts from each 10-degree azimuthal sector. To show the azimuthal variations more clearly, only diffraction from the {111} planes is included.

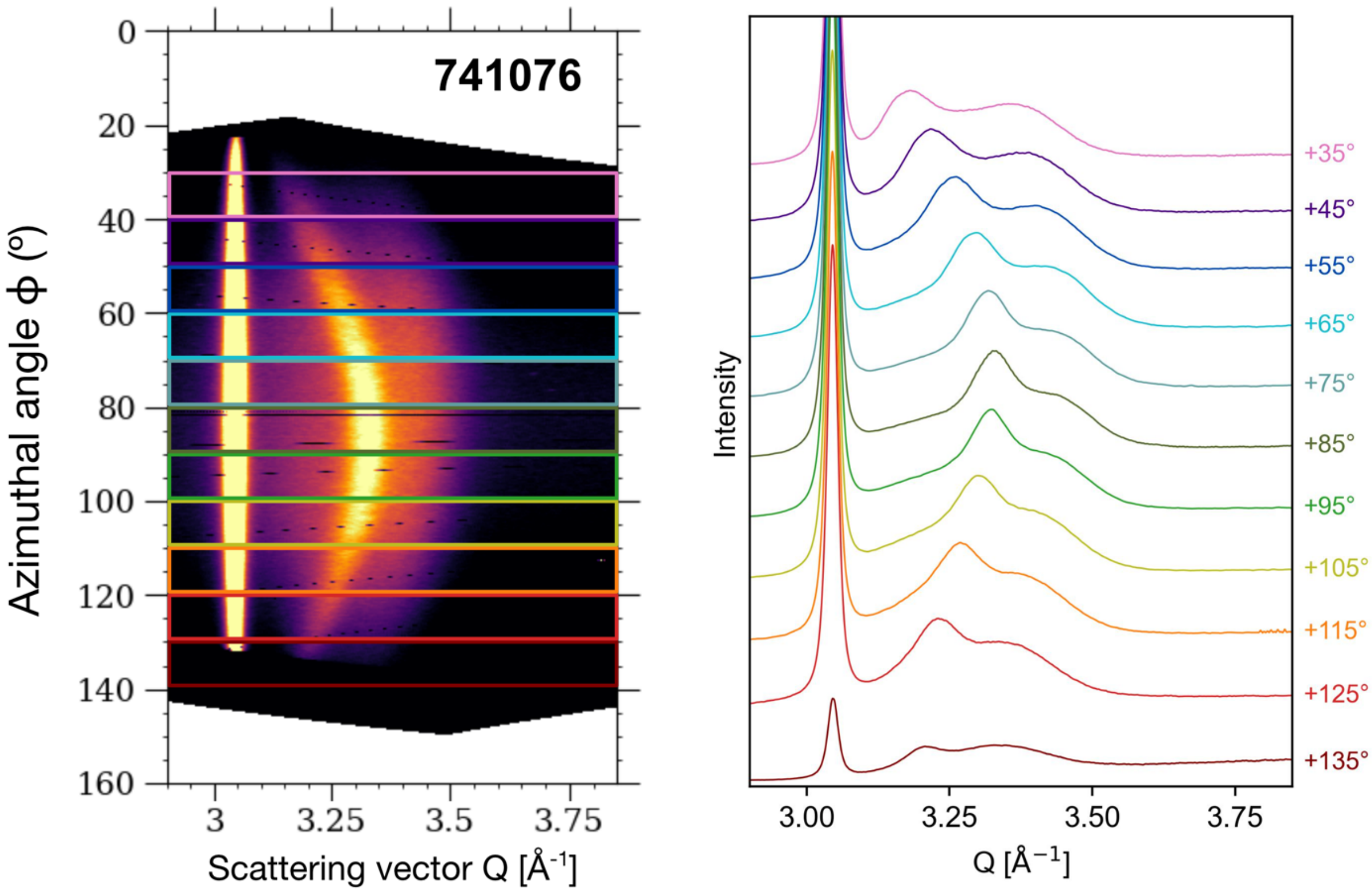

**Fig. S12.** Diffraction pattern of Run 741076 (NPD shocked to 227 ± 8 GPa), with lineouts from each 10-degree azimuthal sector. To show the azimuthal variations more clearly, only diffraction from the {111} planes is included.

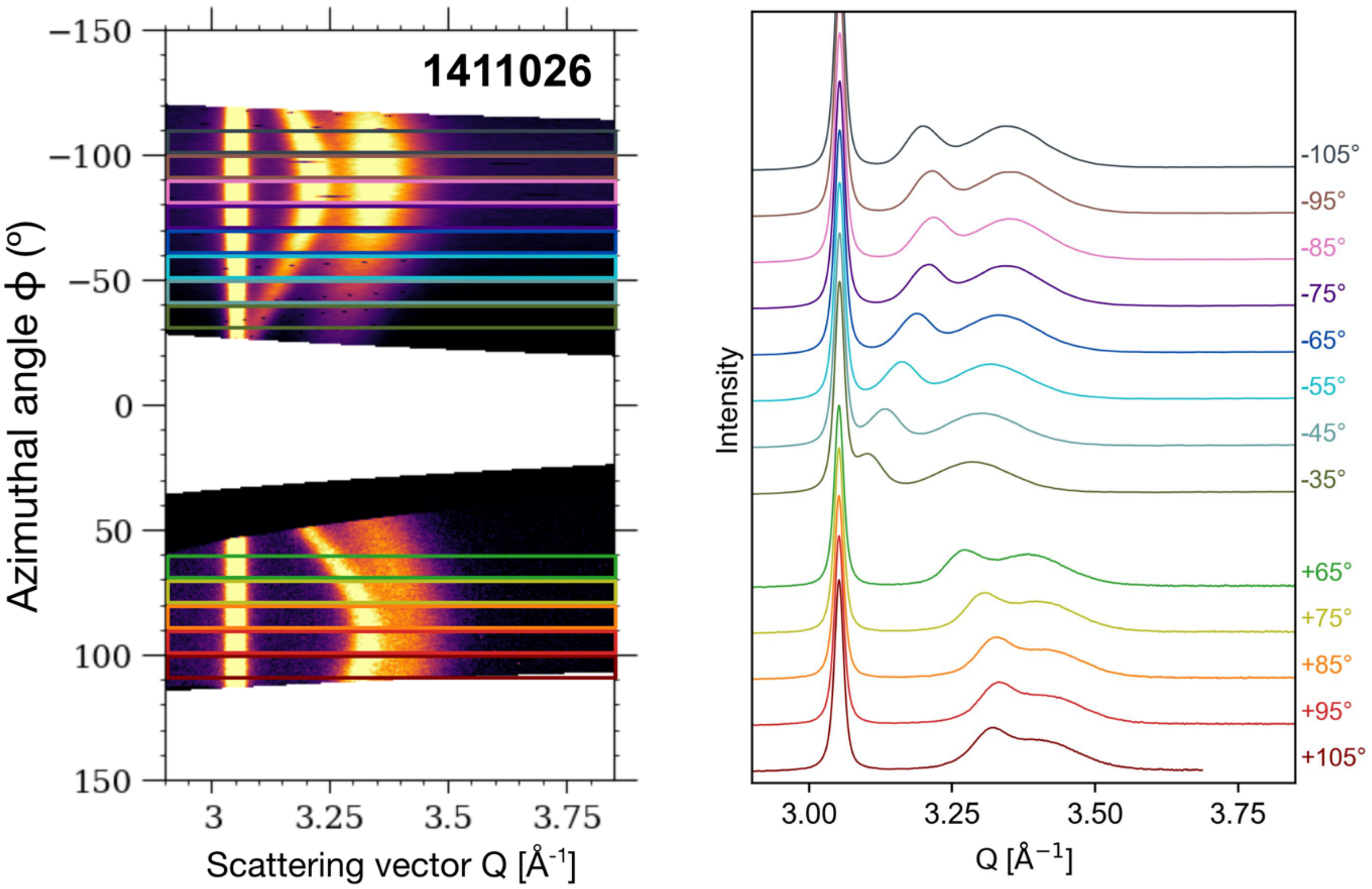

**Fig. S13.** Diffraction pattern of Run 1411026 (NPD shocked to 221 ± 8 GPa), with lineouts from each 10-degree azimuthal sector. To show the azimuthal variations more clearly, only diffraction from the {111} planes is included.

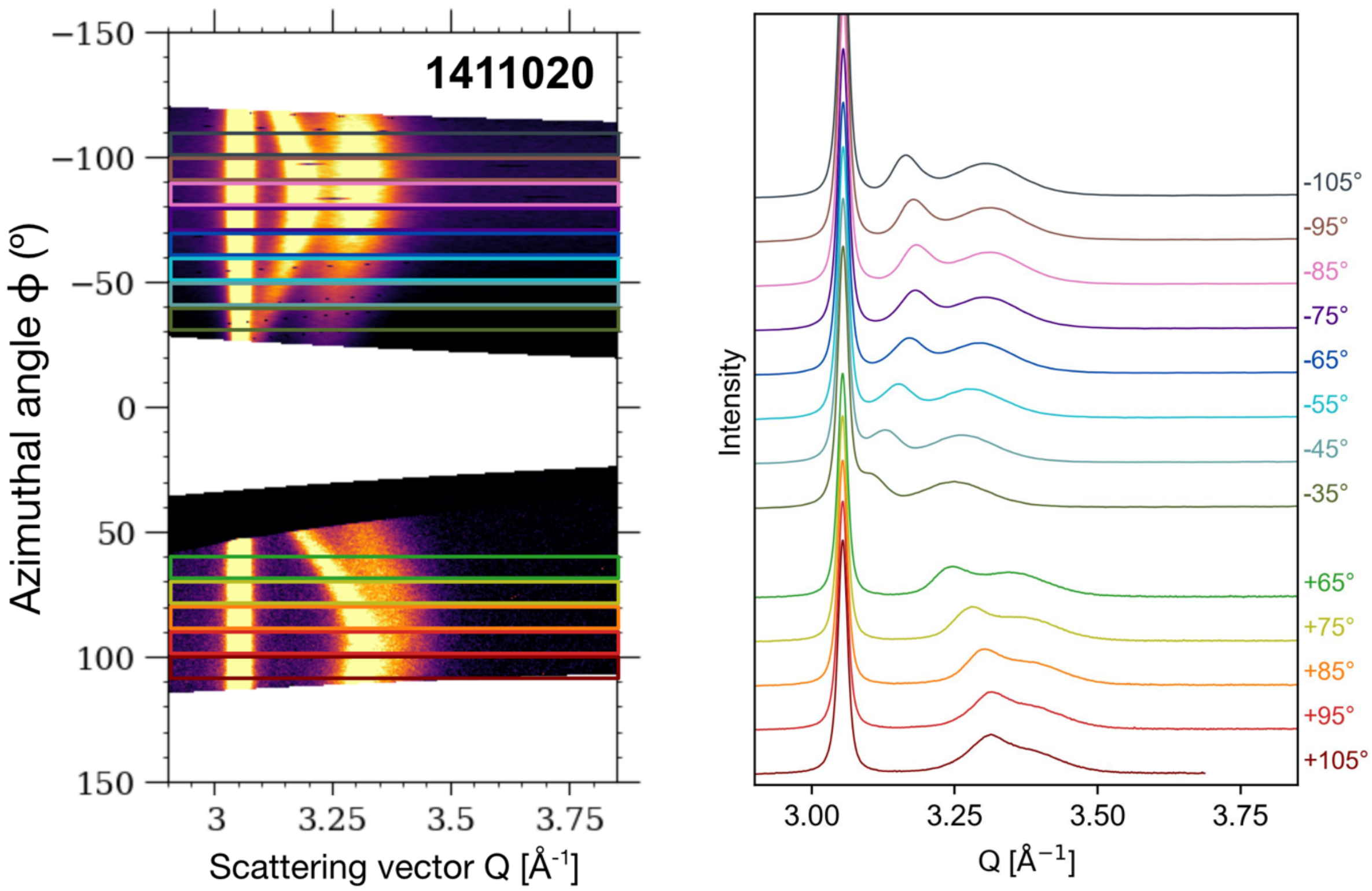

**Fig. S14.** Diffraction pattern of Run 1411020 (NPD shocked to 181 ± 8 GPa), with lineouts from each 10-degree azimuthal sector. To show the azimuthal variations more clearly, only diffraction from the {111} planes is included.

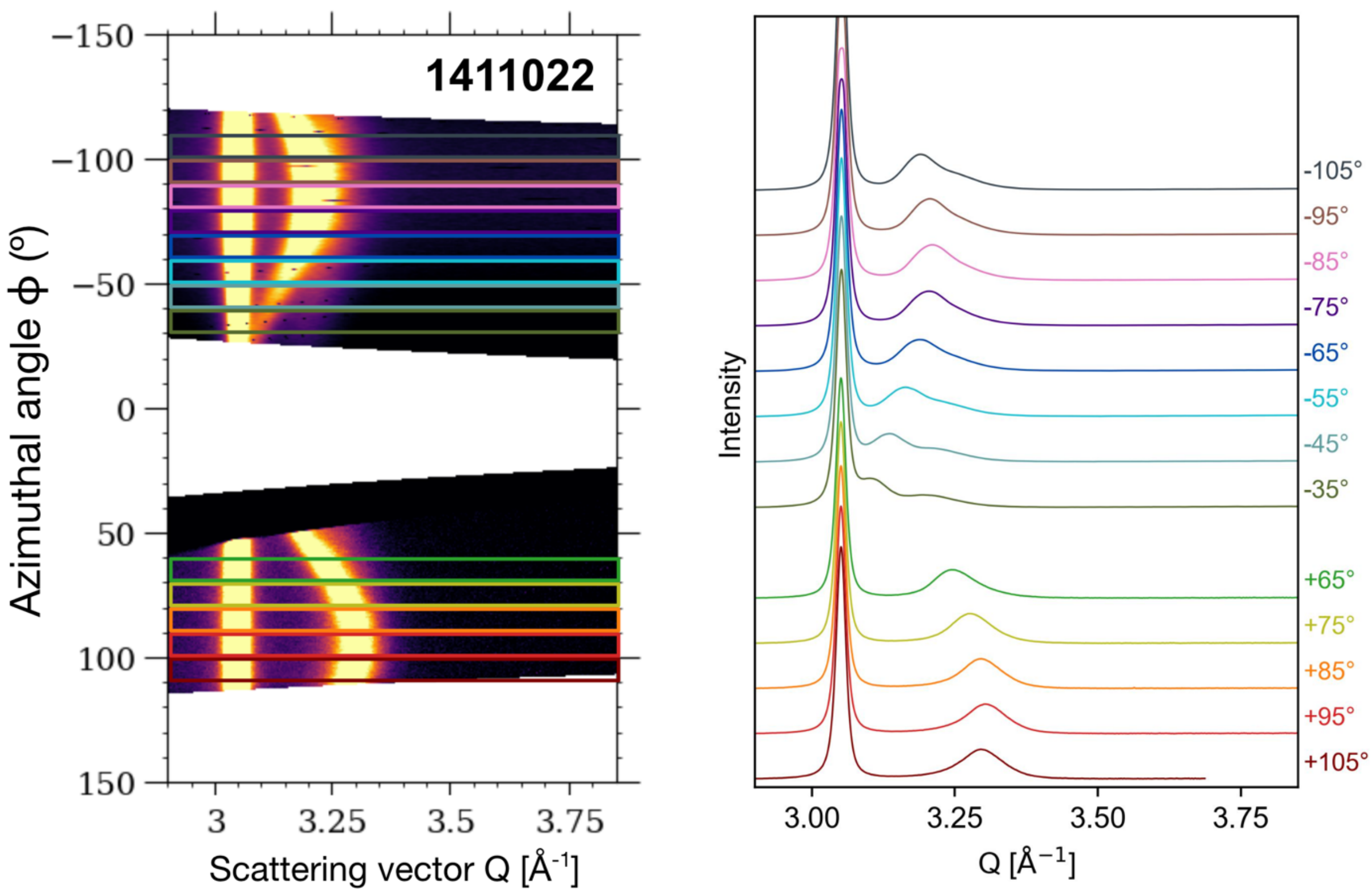

**Fig. S15.** Diffraction pattern of Run 1411022 (NPD shocked to 122 ± 12 GPa), with lineouts from each 10-degree azimuthal sector. To show the azimuthal variations more clearly, only diffraction from the {111} planes is included.

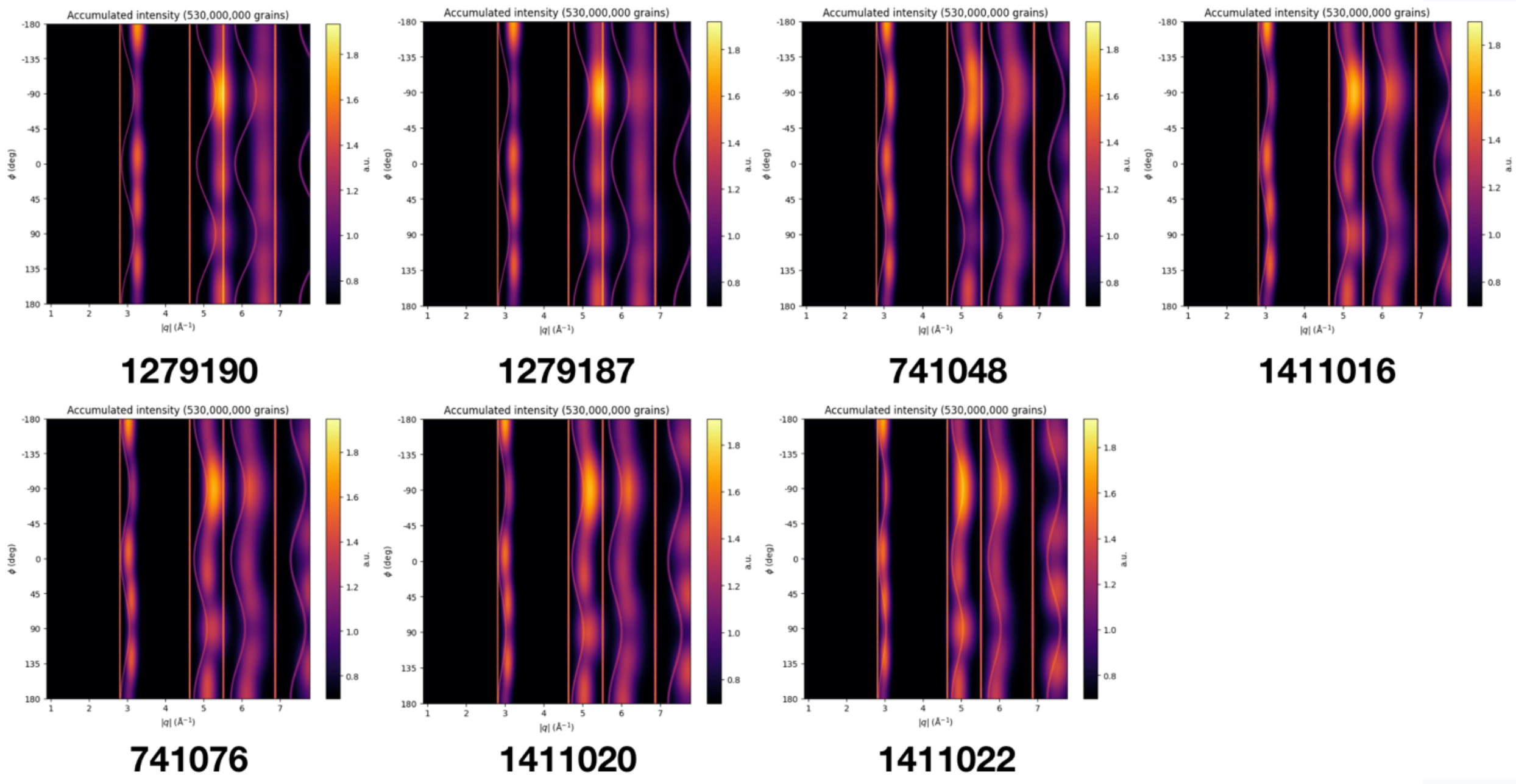

**Fig. S16.** Diffraction simulations of <110> fiber texture from forward model at different beam to shock angles, X-ray energies, and strain states, for direct comparison to each shock experiment. Run 741048 is not included as it is shown in Fig. 3 of the main text. A comparison of the corresponding simulation with each experimental diffraction pattern reveals that the patterns at shock stresses of 425 ± 14 GPa and above are consistent with <110> fiber texture formation.

| Shot ID | Beam to Shock (°) | E (keV) | $\varepsilon^{\ell}_{\parallel}$ | $\varepsilon^{\ell}_{\perp}$ | $\varepsilon^{p}_{\parallel}$ | $\varepsilon^{p}_{\perp}$ | $\varepsilon^{t}_{\parallel}$ | ρ (g/cm³) | $\sigma_{\parallel}$ (GPa) | $2\tau$ (GPa) | $P_M$ (GPa) |
|---|---|---|---|---|---|---|---|---|---|---|---|
| 1279190 | 108 | 10.0 | 0.1402 ± 0.0010 | 0.1396 ± 0.0010 | 0.3116 ± 0.0059 | –0.1396 ± 0.0010 | 0.452 ± 0.006 | 5.52 ± 0.02 | 565 ± 14 | 2 ± 6 | 564 ± 14 |
| 1279187 | 108 | 10.0 | 0.1347 ± 0.0006 | 0.1250 ± 0.0005 | 0.2770 ± 0.0030 | –0.1250 ± 0.0005 | 0.412 ± 0.003 | 5.30 ± 0.01 | 476 ± 10 | 35 ± 3 | 452 ± 10 |
| 741048 | 110 | 10.0 | 0.1343 ± 0.0005 | 0.1141 ± 0.0023 | 0.2522 ± 0.0088 | –0.1141 ± 0.0023 | 0.387 ± 0.009 | 5.17 ± 0.03 | 425 ± 14 | 69 ± 8 | 379 ± 15 |
| 1411016 | 115 | 12.0 | 0.1175 ± 0.0014 | 0.0815 ± 0.0010 | 0.1775 ± 0.0058 | –0.0815 ± 0.0010 | 0.295 ± 0.006 | 4.72 ± 0.02 | 273 ± 9 | 93 ± 4 | 211 ± 9 |
| 741076 | 110 | 10.0 | 0.1133 ± 0.0005 | 0.0684 ± 0.0018 | 0.1487 ± 0.0059 | –0.0684 ± 0.0018 | 0.262 ± 0.006 | 4.57 ± 0.02 | 227 ± 8 | 107 ± 5 | 156 ± 9 |
| 1411026 | 115 | 12.0 | 0.1118 ± 0.0016 | 0.0670 ± 0.0012 | 0.1455 ± 0.0065 | –0.0670 ± 0.0012 | 0.257 ± 0.006 | 4.54 ± 0.02 | 221 ± 8 | 106 ± 5 | 151 ± 9 |
| 1411020 | 115 | 12.0 | 0.0993 ± 0.0020 | 0.0580 ± 0.0015 | 0.1248 ± 0.0077 | –0.0580 ± 0.0015 | 0.224 ± 0.007 | 4.40 ± 0.03 | 181 ± 8 | 90 ± 6 | 121 ± 9 |
| 1411022 | 115 | 12.0 | 0.0761 ± 0.0030 | 0.0436 ± 0.0017 | 0.0922 ± 0.0101 | –0.0436 ± 0.0017 | 0.168 ± 0.010 | 4.16 ± 0.03 | 122 ± 12 | 62 ± 7 | 82 ± 10 |

**Table S1. Table of experimental results.** From left to right: Shot ID, angle from beam direction to shock direction, X-ray energy, longitudinal lattice strain, transverse lattice strain, longitudinal plastic strain, transverse plastic strain, uniaxial total strain, density, longitudinal stress, flow strength and mean pressure. All strains are given as compressive strain, i.e., positive values indicate compression while negative values indicate expansion.

| $\sigma_{\|\|}$ (GPa) | $2\tau$ (GPa) | $P_M$ (GPa) | $\varepsilon^t_{\|\|}$ | $\rho$ (g/cm$^3$) | $\rho_{partial}$ (m$^{-2}$) | $\rho_{full}$ (m$^{-2}$) | Normalized $\rho_{partial}$ (m$^{-2}$) | Normalized $\rho_{full}$ (m$^{-2}$) |
|---|---|---|---|---|---|---|---|---|
| 623 ± 2 | 4 ± 1 | 620 ± 2 | 0.564 ± 0.003 | 5.96 ± 0.01 | 1.68 x $10^{16}$ | 1.39 x $10^{17}$ | 2.43 x $10^{16}$ | 2.00 x $10^{17}$ |
| 484 ± 3 | 20 ± 5 | 470 ± 5 | 0.486 ± 0.002 | 5.51 ± 0.01 | 2.30 x $10^{16}$ | 1.25 x $10^{17}$ | 2.72 x $10^{16}$ | 1.48 $10^{17}$ |
| 366 ± 2 | 66 ± 7 | 322 ± 5 | 0.390 ± 0.006 | 5.01 ± 0.02 | 3.11 x $10^{16}$ | 7.13 x $10^{16}$ | 3.77 x $10^{16}$ | 7.91 x $10^{16}$ |
| 260 ± 4 | 87 ± 9 | 202 ± 7 | 0.290 ± 0.011 | 4.53 ± 0.03 | 3.26 x $10^{16}$ | 3.02 x $10^{16}$ | 3.46 x $10^{16}$ | 3.21 x $10^{16}$ |
| 227 ± 4 | 91 ± 8 | 167 ± 7 | 0.256 ± 0.012 | 4.38 ± 0.04 | 2.91 x $10^{16}$ | 2.20 x $10^{16}$ | 3.04 x $10^{16}$ | 2.30 x $10^{16}$ |
| 196 ± 5 | 90 ± 7 | 136 ± 7 | 0.223 ± 0.013 | 4.24 ± 0.04 | 1.59 x $10^{16}$ | 1.15 x $10^{16}$ | 1.64 x $10^{16}$ | 1.19 x $10^{16}$ |
| 167 ± 4 | 85 ± 5 | 110 ± 5 | 0.193 ± 0.010 | 4.11 ± 0.03 | 1.03 x $10^{16}$ | 8.13 x $10^{15}$ | 1.05 x $10^{16}$ | 8.26 x $10^{15}$ |
| 139 ± 2 | 79 ± 3 | 86 ± 3 | 0.161 ± 0.006 | 3.98 ± 0.02 | 2.80 x $10^{15}$ | 2.84 x $10^{15}$ | 2.83 x $10^{15}$ | 2.87 x $10^{15}$ |
| 108 ± 2 | 67 ± 1 | 63 ± 2 | 0.130 ± 0.002 | 3.86 ± 0.01 | 5.10 x $10^{14}$ | 4.76 x $10^{14}$ | 5.14 x $10^{14}$ | 4.79 x $10^{14}$ |

**Table S2. Table of MD simulation results.** From left to right: longitudinal stress, flow strength, mean pressure, uniaxial total strain, density, a/6<112> partial dislocation density, a/2<110> full dislocation density, a/6<112> partial dislocation density normalized to proportion of cubic diamond atoms and a/2<110> full dislocation density normalized to proportion of cubic diamond atoms. Dislocation densities are calculated for the grain interiors and are normalized to the proportion of cubic diamond atoms. Note that the starting density ($\rho_0$) in the MD simulations is 3.39 g/cm$^3$, lower than that of the experiments. This is due to the smaller grain size (8 nm) used in the simulations, which causes a larger proportion of atoms incorporated in grain boundaries, which are less dense than the grain interiors.